\documentclass[12pt]{article} 

\usepackage{graphicx}
\usepackage{amsmath}
\usepackage{amssymb}
\usepackage{hyperref}
\usepackage[height=8.8in,width=6.45in]{geometry}
\usepackage{tikz}
\usepackage{cite}

\numberwithin{equation}{section}

\newcommand{\cL}{\mathcal{L}}
\newcommand{\cM}{\mathcal{M}}
\newcommand{\cN}{\mathcal{N}}
\newcommand{\cO}{\mathcal{O}}

\newcommand{\beq}{\begin{equation}}
\newcommand{\eeq}{\end{equation}}
\newcommand{\be}{\begin{equation}}
\newcommand{\ee}{\end{equation}}
\newcommand{\bea}{\begin{equation}\begin{aligned}}
\newcommand{\eea}{\end{aligned}\end{equation}}

\begin{document}
\begin{titlepage}

\begin{flushright}

\end{flushright}

\vskip 3cm

\begin{center}
{\Large \bf
Correlation functions in massive Landau-Ginzburg orbifolds and tests of dualities
}

\vskip 2.0cm

Wei Gu

\bigskip
\bigskip

\begin{tabular}{ll}
Center of Mathematical Sciences,
Harvard University, Cambridge, MA 02138, USA\\
\end{tabular}

\vskip 1cm

\textbf{Abstract}
\end{center}

\medskip
\noindent
In this paper we discuss correlation function computations in massive topological Landau-Ginzburg orbifolds, extending old results of Vafa \cite{Vafa:1990mu}.  We then apply these computations to provide further tests of the nonabelian mirrors proposal and two-dimensional Hori-Seiberg dualities with $(S)O_{\pm}$ gauge groups and their mirrors.

\bigskip
\vfill
\end{titlepage}

\setcounter{tocdepth}{2}
\tableofcontents
\section{Introduction}
Gauged linear sigma models (GLSMs)\footnote{In this paper, we only focus on the $\cal{N}$=(2,2) worldsheet supersymmetry.} \cite{Witten:1993yc} were originally motivated by investigating the correspondence between Calabi-Yau and Landau-Ginzburg model conjectured in \cite{Greene:1988ut}. They also play an important role in understanding mirror symmetry. For abelian gauge theories, for example, Morrison and Plesser \cite{Morrison:1995yh, Maxfield:2019czc} interpreted the mirror symmetry between two Calabi-Yaus as a duality between corresponding GLSMs, and Hori and Vafa \cite{Hori:2000kt, Hori:2000ck} constructed Landau-Ginzburg models (LGs) mirror to GLSMs.\\

For abelian GLSMs, many results are known. However, the nonabelian GLSMs are still under active development, although there has been a lot of progress in understanding many aspects of nonabelian GLSMs over the last fifteen years. Following \cite{Witten:1993yc, Witten:1993xi}, Hori and Tong \cite{Hori:2006dk} systematically studied the dynamical behavior of GLSMs with $U(k)$ as well as $SU(k)$ groups and they found two different Calabi-Yau geometries are related by a GLSM, an extension of Witten's CY/LG correspondence. (See also \cite{Caldararu:2007tc} for the similar phenomenon in abelian GLSMs). In 2011 Hori \cite{Hori:2011pd} investigated more nonabelian gauge groups which admit infrared dualities, and these are reminiscent of 4d ${\cal N}$ = 1 Seiberg dualities \cite{Seiberg:1994pq}. Furthermore, these dualities can be tested by studying the exact results obtained from the supersymmetric localization \cite{Benini:2012ui,Doroud:2012xw, Benini:2015noa, Closset:2015rna}, a review on these can be found in \cite{Pestun:2016zxk,Park:2016dpb}.  From the two-dimensional partition functions, the authors of \cite{Halverson:2013eua} observed that each nonabelian GLSM has a corresponding abelian-like GLSM which they called the associated Cartan theory\footnote{One can test this correspondence beyond the partition function; for example, one can study the match of elliptic genera. \cite{Benini:2013nda,Benini:2013xpa,Gadde:2013ftv, Gu:2019rty}}. All these developments inspired the author of this paper with Eric Sharpe to propose the mirror construction for nonabelian GLSMs \cite{Gu:2018fpm}. After that, two followup work were aimed to do more consistency checks to the mirror proposal as well as other physical results in the gauge theories. One is the mirrors of pure gauge theories of exceptional groups studied in \cite{Chen:2018wep} and the other is about the consistency check of Hori-Seiberg dual \cite{Hori:2011pd,Seiberg:1994pq} in the mirror LGs \cite{Gu:2019zkw}. See also \cite{Morrison:1994fr,Melnikov:2005tk,Donagi:2007hi,Melnikov:2010sa,Jockers:2012zr,Jockers:2012dk,Gomis:2012wy,Closset:2014pda,Gomis:2015yaa,Closset:2015ohf,Hori:2016txh,Aharony:2016jki,Gu:2017nye ,CaboBizet:2017fzc,Knapp:2019cih,Gu:2019byn} for some related works on these directions.\\

In this paper, we study B-twisted Landau-Ginzburg models with massive orbifolds following the discussion in \cite{Gu:2019zkw}. We provide formulas for correlation functions. A parallel discussion in the A-twisted gauge theories can be found in \cite{Closset:2017vvl}, which provides some pertinent tools. The motivation for \cite{Closset:2017vvl} was to test Hori-Seiberg dualities in 2d gauge theories using the correlation function obtained with supersymmetric localization. Our motivation for this paper is similar: to test the Hori-Seiberg dualities by comparing the correlation function in B-twisted mirror LG models. Thus, we will derive the correlation functions of the B-twisted LGs with massive orbifold as well as give a physical explanation of assigning different values of weights for different orbifolds in summing the twisted sectors. In section \ref{TLGMMO}, we provide the correlation function for the LGs with massive orbifolds. One can find it is much more straightforward than the gauge theory side to obtain the correlation functions of massive orbifolds. In section \ref{HSMS}, we summarize some basic facts of the ${\cal N}$=(2,2) gauge theories as well as the nonabelian mirror proposals, these are for serving the following sections. Finally, in section \ref{SOOD} and \ref{OOD} we use the formula of the correlation function of massive orbifolds to check the match of correlation function across the Hori-Seiberg dualities in the mirrors.

\section{Orbifolds of topological Landau-Ginzburg models}\label{TLGMMO}
In this section, we provide a formula for correlation functions of topological Landau-Ginzburg models with massive orbifolds. It is an extension of Vafa's work \cite{Vafa:1990mu}, and most techniques we use here have already been developed in that paper.\\

The Landau-Ginzburg theories we mainly consider in this paper are defined on $\mathbb{Z}_{2}$ orbifolds of noncompact Calabi-Yau manifolds ${\cM}$s with superpotentials, where the 2d worldsheet ${\cN}=(2,2)$ theories are still B-twistable even if the $\mathbb{Z}_{2}$ breaks\footnote{Our definition of Calabi-Yau manifold in this paper is: existing a nowhere vanishing holomorphic volume form.}  the Calabi-Yau structure \cite{Sharpe:2006qd}. If the superpotential is a quasi-homogeneous function, we can apply A-twist to this model as well.\\

For the flat worldsheet, one can formally write out the Lagrangian in terms of superspace coordinates as
\begin{equation}\label{LLG}
  {\cL}=\int d^{4}\theta K(\overline{X}_{i}, X_{i})+\int d^{2}\theta W\left(X_{i}\right)+c.c.
\end{equation}
The model may have some discrete global symmetry $G_{1}$ as well as the discrete gauge symmetry $G_{2}$. The K\"{a}hler potential and the superpotential should be invariant\footnote{Or up to a 2$\pi$i shift of the theta angle; for example, see the mirror of the pure SO(3) gauge theory \cite{Gu:2018fpm}.} under the transformation of these groups. In this paper, we are interested in the B-twisted topological Landau-Ginzburg model which can be defined on an arbitrary curved closed worldsheet-$\Sigma_{g}$. The action, in terms of component fields, is

\begin{eqnarray}\label{LLG2}\nonumber
   S&=&\int d^{2}z\left(\partial_{\mu}x^{i}\partial^{\mu}x^{\bar{j}}\eta_{i\bar{j}}-i\eta_{i\bar{j}}\psi^{\bar{j}}\partial_{\bar{z}}\rho^{i}_{z} \right. \\\nonumber
   &&\left. +i\eta_{i\bar{j}}\overline{\psi}^{\bar{j}}\partial_{z}\rho^{i}_{\bar{z}}+\frac{1}{8}\eta^{\bar{j}i}\partial_{\bar{j}}\overline{W}\partial_{i}W  \right.\\\nonumber
   &&\left.+\frac{1}{4}\left(\partial_{i}\partial_{j}W\rho^{i}_{\bar{z}}\rho^{j}_{z}\right)+\frac{1}{4}\left(\partial_{\bar{i}}\partial_{\bar{j}}\overline{W}\bar{\psi}^{\bar{i}}\bar{\psi}^{\bar{j}}\right)  \right).
\end{eqnarray}
The B-twisted supersymmetric transformation of these component fields are
\begin{eqnarray}\label{ST}\nonumber
  \delta x^{i} = 0, \quad&&\quad \delta\overline{x}^{\bar{i}}=-\overline{\epsilon}\left(\psi^{\bar{i}}+\overline{\psi}^{\bar{i}}\right) \\\nonumber
  \delta(\psi^{\bar{i}}-\overline{\psi}^{\bar{i}}) =\bar{\epsilon}\eta^{\bar{i}j}\partial_{j}W, \quad&&\quad \delta\left(\psi^{\bar{i}}+\overline{\psi}^{\bar{i}}\right)=0,  \\\nonumber
   \delta\rho^{i}_{\mu}&=&-2\bar{\epsilon}\partial_{\mu}x^{i}.
\end{eqnarray}
The $x$ and $\psi$ fields are worldsheet scalars while the $\rho$s are the worldsheet one-form fields. One can easily find that the physical observables are holomorphic functions of $x^{i}$ modulo the $\partial_{i}W$ which is $Q_{B}$ exact. The chiral ring is simply the ring of functions on $\cM$ module the ideal of functions of the form $\partial_{i}W$. If the theory has gauge group $G_{2}$, all these physical observables should be gauge invariant under this gauge symmetry. However, we do not require the physical operators to be invariant under the global symmetry $G_{1}$, they can be some nontrivial representation of the global symmetry.\\

Now, we (re)-derive the correlation functions of B-twisted Landau-Ginzburg models with some discrete symmetries. Most cases have already been derived by Vafa \cite{Vafa:1990mu}, with one special case missed in Vafa's paper. We fill this gap in the literature and will also give comments on the existence formulas of the correlation functions of topological Landau-Ginzburg with massless orbifolds. The path integral of the correlation function formally can be written as
\begin{equation}\label{PI}
  \langle \cO \rangle=\int \prod_{i}[Dx_{i}][D\overline{x}_{i}][D\psi_{i}][D\overline{\psi}_{i}][D\rho_{i}][D\overline{\rho}_{i}]e^{-S}\cO.
\end{equation}
We can apply localization techniques to compute the path integral, and the path integral depends solely on the $Q_{B}$ fixed points obeying
\begin{equation}\label{QFS}
  \partial_{\mu}x=0,\quad\quad \partial_{i}W=0,
\end{equation}
so only constant maps into critical points of $W$ are important to the path integral. We should emphasize that the discrete gauge symmetry might permute the variables, and the critical loci should modulo these permutations. We also assume the critical loci are isolated. For the non-isolated case, we could possibly turn on the RG-flow relevant term to make it isolated. One exception is the massless orbifold theory which we comment later on. All these were already discussed in the literature, except one special case was missed. Imagine we have the following vacua equation
\begin{equation}\label{VCL}
  (x-a_{1})\cdots(x-a_{n})=(-1)^{n-1}(-x-a_{1})\cdots(-x-a_{n}), \quad a_{i}\neq a_{j} \quad {\rm if} \quad i \neq j.
\end{equation}
From the above, one can notice the theory could have a $\mathbb{Z}_{2}$  discrete symmetry acting on the variable as
\begin{equation}\label{Z2O}
  x\mapsto -x.
\end{equation}
If one $x_{m}$ is vanishing, there is a special vacuum: $x=0$. If the target space is an orbifold, the vacuum $x=0$ intersects with the orbifold fixed point and the twisted sector of this orbifold has a non-vanish Hessian factor that can contribute to the high genus correlation function. We call this orbifold in this paper as $massive$ $orbifold$ and we should take into account the contribution from the twisted sector of this orbifold \cite{Dixon:1985jw,Dixon:1986jc}. This $\mathbb{Z}_{2}$ could be a subgroup of a bigger discrete group $G_{2}$ which will be mainly discussed in the later application. One can consider tensor product of massive orbifolds such as $\mathbb{Z}^{N}_{2}$. This may be useful for studying the massive mirror-pair LG theories.\\

Now, we can perform the path integral if we know the constant modes of each field. For Landau-Ginzburg models without the massive $\mathbb{Z}_{2}$ orbifold, the constant modes are well known in the literature. Each of $x^{i}$, $\overline{x}^{\bar{i}}$, $\psi^{i}$ and $\overline{\psi}^{\bar{i}}$ has a single constant mode. By contrast, for the genus-$g$ worldsheet, each of one-form fields $\rho^{i}_{z}$ and $\rho^{i}_{\bar{z}}$ has $g$ ``constant modes." Now, we can evaluate the path integrals for each constant modes:
\begin{equation}\label{C1I}
  \int\prod_{i}dx_{i}d\overline{x}_{\bar{i}}e^{-\frac{1}{4}\eta^{i\bar{j}}\partial_{i}W\partial_{\bar{j}}\overline{W}}=\sum_{\rm{vacua}}\frac{1}{\overline{H}H},
\end{equation}
\begin{equation}\label{C2I}
  \int\prod_{i}d\psi_{i}d\overline{\psi}_{\bar{i}}e^{-\frac{1}{2}\partial_{\bar{i}}\partial_{\bar{j}}\overline{W}\psi^{\bar{i}}\overline{\psi}^{\bar{j}}}=\sum_{\rm{vacua}}\overline{H},
\end{equation}
\begin{equation}\label{C3I}
  \int\prod_{i}d\rho^{g}_{i}d\overline{\rho}^{g}_{\bar{i}}e^{-\frac{1}{2}\partial_{i}\partial_{j}W\rho^{i}_{\bar{z}}\rho^{j}_{z}}=\sum_{\rm{vacua}}H^{g},
\end{equation}
where $H$ is
\begin{equation}\label{Hessian}
  H=\det\partial_{i}\partial_{j}W.
\end{equation}
Therefore, the general expression of correlation functions is
\begin{equation}\label{CL}
  \langle \cO \rangle_{g}=\sum_{\rm{vacua}}\frac{\cO}{H^{1-g}}.
\end{equation}

Eq.(\ref{CL}) has already been derived by Vafa \cite{Vafa:1990mu}. Furthermore, he also computed the correlation functions of SCFTs. For the case with massive orbifolds, the twisted sector contributes to the higher genus correlation functions due to the ``constant modes" of one-form fields $\rho$ in the twisted sector. Therefore, the derivation should be refined for this orbifold case as well. More precisely, for $\mathbb{Z}_{2}$ orbifold fixed point twisted sector, the fields obey the following antiperiodic condition
\begin{equation}\label{BC}
  X(\sigma+2\pi)=-X(\sigma).
\end{equation}
So the worldsheet scalar fields $x$, $\overline{x}$, $\psi^{i}$ and $\overline{\psi}^{\bar{i}}$ all have no constant mode. But we still have $g-1$ for each worldsheet one-form field \cite{hatcher2002algebraic}. The path integral of the twisted sectors are surprising much easier than the untwisted sector\footnote{In gauge theory side, the derivation of the correlation function from the twisted sector is more complicated than the untwisted sector \cite{Closset:2017vvl}.}, and it is
  \begin{equation}\label{tC3I}
  \int\prod_{i}d\rho^{g-1}_{i}d\overline{\rho}^{g-1}_{\bar{i}}e^{-\frac{1}{2}\partial_{i}\partial_{j}W\rho^{i}_{\bar{z}}\rho^{j}_{z}}=\sum_{\rm{twisted}}H^{g-1}(x=0).
\end{equation}
so the twisted sector's piece to the correlation function is
\begin{equation}\label{TCL}
\langle\cO\rangle_{\rm{twisted}}=\sum_{\rm{twisted}}\frac{\cO}{H^{1-g}(x=0)}.
\end{equation}
One can easily find that it shares a similar formula with the untwisted sector's at the vacua $x=0$. One may notice that we have already set all the $2g$ cycles of the genus-g worldsheet $\Sigma_{g}$ with the antiperiodic boundary condition in getting the formula (\ref{tC3I}). However, we do not have to impose all of the cycles to have the antiperiodic boundary condition. The situation where periodic boundary condition on some cycles while antiperiodic boundary condition on the others also belongs to the twisted sector, and one can easily derive that the path integral of such is also expressed in terms of formula (\ref{TCL}). The untwisted sector corresponds to the field that has the periodic boundary condition on all of the worldsheet cycles. The path integral requires us to sum all of the possibilities with weight ${\cal}{v}$ for the untwisted sector and ${\cal}{w}$ for the twisted sector to manifest the different orbifold projections. These weights are related to the $discrete$ $torsion$ \cite{Vafa:1986wx,Sharpe:2000ki}. Thus, one may suggest that all these weights should all be 1 because the discrete torsion ($H^{2}\left(\mathbb{Z}_{2}, U(1)\right)$) is trivial . However, as mentioned in \cite{Hori:2011pd} that dressing $\mathbb{Z}_{2}$ with the operator $(-1)^{F}$ would affect the orbifold projection. Following Hori's normalization convention of the operator $(-)^{F}$ on the Ramond ground state, the weights may have two possible values: 1 and -1. One can figure out the values of these two weights by studying the genus-zero as well as the genus-one correlation functions. A concrete example would be useful for understanding all these details, so let us first consider the simplest massive Landau-Ginzburg model: a target space $\mathbb{C}$ which we denote $X$, with the superpotential,
\begin{equation}\label{}\nonumber
  W=X^{2}
\end{equation}
The theory has a $\mathbb{Z}_{2}$ global symmetry such that the Lagrangian is invariant under the transformation $X\mapsto -X$. The ground state $dW=0$ and the Hessian $d^{2}W_{\rm vacua}$ can be computed:
\begin{equation}\label{}\nonumber
  X=0\quad\quad\quad d^{2}W_{\rm vacua}=2.
\end{equation}
One can compute the correlation function of the genus-g worldsheet topological B-twisted LG theory
\begin{equation}\label{}\nonumber
  \langle{\cal O}(X)\rangle_{g}={\cal O}(X)\left(d^{2}W\right)^{^{g-1}}_{\rm vacua}.
\end{equation}
The question is what happens if we gauge the $\mathbb{Z}_{2}$ global symmetry. It is known that the theory after gauging the $\mathbb{Z}_{2}$ global symmetry is dual to the theory before gauging, so we only have one ground state at $X=0$. Let us consider the genus-zero correlation function first,
\begin{equation}\label{}\nonumber
  \langle {\cal O}(X)\rangle_{g=0}=\frac{1}{|\mathbb{Z}_{2}|}\frac{{\cal O}(X)}{d^{2}W}_{\rm vacua}=\frac{{\cal O}(X)}{2d^{2}W}_{\rm vacua}.
\end{equation}
So the genus zero correlation function tells us how the Hessian rescales when we gauged some discrete global symmetry. Next, from the torus correlation function which we have to take into account the twisted sectors, we find
\begin{equation}\label{}\nonumber
   \langle {\cal O}(X)\rangle_{g=1}=\frac{1}{2}{\cal}{v}\langle{\cal O}(X)\rangle_{\rm untwisted}+\frac{3}{2}{\cal}{w}\langle{\cal O}(X)\rangle_{\rm twisted}=\frac{{\cal}{v}+3{\cal}{w}}{2}{\cal O}(X)_{X=0}.
\end{equation}
The factor three in front of $w$ can be explained by the fact that the genus-one curve has three orbifold twisted sectors from the possible anti-boundary conditions on the two cycles of the torus. We also have used the fact that the Hessian of the twisted sector equals to the Hessian of the untwisted sector. Because the number of the ground state is one, we thus impose
\begin{equation}\label{}\nonumber
  \frac{{\cal}{v}+3{\cal}{w}}{2}=1.
\end{equation}
This suggests that
\begin{equation}\label{2op}
  {\cal}{v}=-1,\quad\quad\quad {\cal}{w}=1.
\end{equation}
One can then use these values of weights for computing the higher genus correlation functions.\\

The next simplest LG model we consider is the target space $\mathbb{C}^{2}/\mathbb{Z}_{2}$ with a superpotential
\begin{equation}\label{}\nonumber
  W=X^{2}_{1}+X^{2}_{2},
\end{equation}
where the orbifold acts on the fields by the following
\begin{equation}\label{}\nonumber
  X_{1}\mapsto -X_{1},\quad\quad\quad X_{2}\mapsto -X_{2}.
\end{equation}
This LG model has two ground states and one of them corresponds to the twisted sector of the orbifold fixed point. Let us again consider the sphere correlation function:
\begin{equation}\label{}\nonumber
  \langle {\cal}{O}(X)\rangle_{g=0}=\frac{1}{|\mathbb{Z}_{2}|}\frac{{\cal}{O}(X)}{d^{2}W}_{\rm vacua}=2\frac{{\cal}{O}(X)}{4d^{2}W}_{X_{i}=0}.
\end{equation}
The factor two in the correlation function indicates that we have taken into account the twisted sector vacua of the $\mathbb{Z}_{2}$ orbifold, so one can read how the Hessian rescales from the genus-zero correlation function. Next, we check the torus correlation function to see what are the weights of this case:
\begin{equation}\label{}\nonumber
  \langle {\cal O}(X)\rangle_{g=1}=\frac{1}{2}{\cal}{v}\langle{\cal O}(X\rangle)_{\rm untwisted}+\frac{3}{2}{\cal}{w}\langle{\cal O}(X)\rangle_{\rm twisted}=\frac{{\cal}{v}+3{\cal}{w}}{2}{\cal O}(X)_{X=0}=2{\cal O}(X)_{X=0}.
\end{equation}
We have picked up the normalization factor two to indicate that we have included the twisted vacua. So one can easily see that we shall impose
\begin{equation}\label{1op}
   {\cal}{v}=1,\quad\quad\quad {\cal}{w}=1.
\end{equation}
We will use these normalization rules later on to check the match of the correlation functions between the dual theories. One can easily notice that the discussion of the above two cases can be generalized to other massive theories. \\

Eventually, we arrive at the genus-g correlation functions of the $\mathbb{Z}_{2}$ orbifold theories
\begin{eqnarray}\label{GZ2CL}\nonumber
  \langle {\cal O}(X)\rangle_{g} &=& \frac{1}{2}\langle {\cal O}(X)\rangle_{X\neq 0}+\frac{1}{2} {\cal}{v}^{g}\langle {\cal O}(X)\rangle_{\rm untwisted}+\frac{1}{2} \left[\left({\cal}{v}+3{\cal}{w}\right)^{g}-{\cal}{v}^{g}\right]\langle {\cal O}(X)\rangle_{\rm twisted} \\
  &=& \frac{1}{2}\langle{\cal O}(X)\rangle_{X\neq 0}+\frac{1}{2}\left({\cal}{v}+3{\cal}{w}\right)^{g} \langle{\cal O}(X)\rangle_{X=0}.
\end{eqnarray}
\\
Our formula is slightly different from the formula derived in \cite{Closset:2017vvl}. For the vacua do not intersect with the orbifold fixed point, we simply divide by two in front of the computation as no twisted sector needs to be summed, while for the vacua intersect with the orbifold fixed point, we assign the weight ${\cal}{v}$ to the untwisted sector while the weight ${\cal}{w}$ to the twisted sector to manifest the different orbifold projections.\\

\noindent\textbf{Generalization}: one may be interested in the case of $\mathbb{Z}^{N}_{2}$ massive LG orbifold such as
\begin{equation}\label{}
  W=\frac{X^{2}_{11}+\cdots+X^{2}_{1i_{1}}}{\mathbb{Z}_{2}}+\cdots+\frac{X^{2}_{N1}+\cdots+X^{2}_{Ni_{N}}}{\mathbb{Z}_{2}}.
\end{equation}
The correlation function can be derived by using the factorization feature of the corresponding path integral
\begin{eqnarray}
  \langle{\cal O}(X)\rangle_{g}&=& \langle{\cal O}_{1}(X)\rangle_{g}\cdots\langle{\cal O}_{N}(X)\rangle_{g} \\\nonumber
   &=&\left(\frac{1}{2} \left({\cal}{v}_{1}+3{\cal}{w}_{1}\right)^{g}\langle {\cal O}_{1}(X)\rangle_{X_{1}=0}\right)\times\\\nonumber
   &\cdots& \times \left(\frac{1}{2} \left({\cal}{v}_{N}+3{\cal}{w}_{N}\right)^{g}\langle {\cal O}_{N}(X)\rangle_{X_{N}=0}\right) \\\nonumber
   &=&\frac{1}{2^{N}}\left({\cal}{v}_{1}+3{\cal}{w}_{1}\right)^{g} \cdots \left({\cal}{v}_{N}+3{\cal}{w}_{N}\right)^{g} \langle{\cal O}(X)\rangle_{X=0}.
\end{eqnarray}\\

\noindent{\textbf{Comment on the massless orbifold}}: for an orbifold LG, the massless orbifold is defined by
\begin{equation}\label{}\nonumber
  d^{2}W=0 \quad\text{at the orbifold vacua}.
\end{equation}
The correlation function of the higher genus ($g>1$) can be obtained by summing twisted sectors:
\begin{equation}\label{}\nonumber
  \langle{\cal O}(X)\rangle\propto f({\cal}{v},{\cal}{w}){\cal O}(X)H^{g-1}_{\rm vacua}=0.
\end{equation}
The higher genus correlation functions of the massless orbifold are vanishing, while the only nontrivial correlation function is the genus-zero correlation function which has already been computed in \cite{Vafa:1990mu}. A similar selection rule in the nonlinear sigma model can be found at \cite{Witten:1988xj}.

\section{Two-dimensional Hori-Seiberg dual and mirror symmetry}\label{HSMS}
In this section, we briefly summarize some features of two-dimensional nonabelian gauge theories and their mirrors. Consider the GLSM with rank $N_{c}$ compact Lie gauge group $\textbf{G}$. It has a vector superfield in adjoint representation of the gauge group $\textbf{G}$ with roots $\alpha^{a}_{\mu}$ as well as it could also have matter fields in the representations of the gauge group with weights $\rho^{a}_{i}$. When matter fields and $W$-bosons are heavy, we can integrate out the matter fields to the ``Coulomb branch,"  to induce a twisted effective superpotential
\begin{equation}\label{TES}
  W_{eff}=-\sum^{N_{c}}_{a=1}t_{a}\Sigma_{a}-\sum_{i,a}\left(\sum_{a}\rho^{a}_{i}\Sigma_{a}-m_{i}\right)\left[\log\left(\sum_{a}\rho^{a}_{i}\Sigma_{a}-m_{i}\right)-1\right]-\sum_{\rm{pos'}}i\pi \alpha^{a}_{\tilde{\mu}}\Sigma_{a},
\end{equation}
where
\begin{equation}\label{}\nonumber
  t_{a}=r_{a}-i\theta_{a}.
\end{equation}
As the $W$-bosons are heavy, we have to impose
\begin{equation}\label{ECL}
  \sigma_{a}\neq \sigma_{b}, \quad\quad\quad {\rm if} \quad\quad a\neq b.
\end{equation}
The $\sigma_{a}$ is the lowest component of the field strength superfield $\Sigma_{a}$. The observables are some gauge invariant function of $\sigma_{a}$ denoted by ${\cal O}(\sigma_{a})$.  The vacua are the solutions of the vacua equation \cite{Nekrasov:2014xaa}:
\begin{equation}\label{VE11}
  \exp\left(\frac{\partial W_{eff}}{\partial \sigma_{a}}\right)=1,\quad\quad\quad\quad\quad \omega\cdot\sigma\neq\sigma,\quad\quad \forall \omega\in W_{\rm \textbf{G}},
\end{equation}
the index $a=1,\cdots, N_{c}$, and the vacua are defined by modulo the Weyl orbifold $W_{\rm \textbf{G}}$. The second condition in (\ref{VE11}) means we have no vacua in the strong coupling region which can be understood from the mirror symmetry \cite{Gu:2018fpm}. \\

When the GLSM couples to the curved spacetime, it is useful to test the dualities by defining a constant term in the path integral \cite{Closset:2017vvl} in keeping track of the overall normalization factor across the duality:
\begin{equation}\label{CCC}
  e^{-\int d^{2}x\sqrt{g}c}=e^{-(g-1)c}=C^{g-1}.
\end{equation}
\\
Furthermore, the flavor symmetry of the theory can be gauged by turning on the twisted masses and the ``flux operators" defined in \cite{Closset:2017vvl}:
\begin{equation}\label{}\nonumber
  \Pi_{F}=\exp\left(\frac{\partial W_{eff}}{\partial m_{F}}\right).
\end{equation}
In this paper, for simplicity, we will not turn on flux operators in the following discussions. However, we will comment on how these operators affect mirror construction later on. For the massive theory with a connected gauge group, the correlation function has a neat formula obtained by the supersymmetric localization or the Bethe ansatz:
\begin{equation}\label{MCF}
  \langle{\cal O}(\sigma)\rangle=\sum_{\rm vacua}\frac{{\cal O}(\sigma)}{{\cal H}^{1-g}(\sigma)},
\end{equation}
where the ${\cal H}$ is
\begin{equation}\label{GH}
  {\cal H}(\sigma)=C\det\left(-\partial_{a}\partial_{b}W_{eff}\right)\left(\prod_{i}\left(\rho^{a}_{i}\sigma_{a}+m_{i}\right)\prod_{\mu}\alpha^{a}_{\mu}\sigma_{a}\right)^{-1}.
\end{equation}
For the disconnected gauge theory like $O(k)$ case, the formula (\ref{GH}) is modified by including the twisted sector from the $\mathbb{Z}_{2}$ orbifold \cite{Closset:2017vvl}. In the next section, we derive the similar results in \cite{Closset:2017vvl} by considering the mirror LG models.\\

Like the higher dimensional supersymmetric gauge theories, the dynamics of 2d gauge theory is very rich. It was proposed by Hori \cite{Hori:2011pd} that two-dimensional gauge theories have infrared-dual pairs which is reminiscent of four-dimensional Seiberg dualities. We briefly summarize some of them here, for $N \geq k$, there exist IR dualities:
\begin{eqnarray}
SO(k) & \leftrightarrow & O_+(N-k+1), \\
O_+(k) & \leftrightarrow & SO(N-k+1), \\
O_-(k) & \leftrightarrow &  O_-(N-k+1),
\end{eqnarray}
where in each case
\begin{itemize}
\item the theory on the left has
$N$ massless vectors $x_1, \cdots, x_N$ ,
with twisted masses $-m_i$ and vanishing R-charges,
and
\item
the theory on the right has $N$ vectors
$\tilde{x}^1, \cdots, \tilde{x}^N$ with all R-charges 1, of twisted masses $m_i$,
along with
$(1/2)N(N+1)$ singlets $s_{ij} = + s_{ji}$ with vanishing R-charges,
$1 \leq i, j \leq N$, of twisted mass $-m_i - m_j$,
and a superpotential
\begin{equation}
W \: = \: \sum_{i,j} s_{ij} \tilde{x}^i \cdot \tilde{x}^j.
\end{equation}
\end{itemize}
The mesons in the two theories are related by
\begin{equation}
s_{ij} \: = \: x_i \cdot x_j.
\end{equation}
The dualities are only claimed to exist when all the theories in question are regular, in the sense of \cite{Hori:2011pd}, which constrains the discrete theta angle.  In \cite{Gu:2019zkw}, we study the mirrors of these gauge theories and we examined the mirrors to each side of the first two dualities above, using an extension of the nonabelian mirrors proposal of \cite{Gu:2018fpm}, and argue that the mirrors of the duals are equivalent.\\

\noindent{\textbf{Nonabelian mirrors}}:
In \cite{Gu:2018fpm}, we proposed that the mirror of a nonabelian GLSM is a Weyl-group-orbifold of a Landau-Ginzburg model. The Landau-Ginzburg model has the following matter fields:
\begin{itemize}
\item  $r$ chiral superfields $\Sigma_{a}$, corresponding to a choice of Cartan subalgebra of the
Lie algebra of $G$,

\item  $N$  chiral superfields $Y_{i}$, each of imaginary periodicity 2$\pi$i, corresponding to matters with arbitrary representations $\mathfrak{R}_m$ in gauge theory,

\item  $L$ chiral superfields $X_{\mu}$, corresponding to the roots of the group $G$

\end{itemize}
with superpotential
\begin{align}\label{MLGS}
  W = & \sum^{r}_{a=1}\Sigma_{a}\left(\sum_{m}\sum_{\rho\in\Lambda_{\mathfrak{R}_m}}\rho^{a}Y_{\rho^m}-\sum^{L}_{\mu=1}\alpha^{a}_{\mu}\ln X_{\mu}-t_{a}\right) \\\nonumber
  & +\sum^{N}_{i=1}m_{i}Y_{i}+\sum^{N}_{i=1}\exp\left(-Y_{i}\right)+\sum^{L}_{\mu=1}X_{\mu}.
\end{align}
The mirror above assumes that the gauge theories do not turn on the ``flavor $flux$ operators." However, if turn on the ``flavor $flux$ operators" in the gauge theories, one needs to redefine the fundamental variables of the mirror to reflect the ``gauging flavor symmetry." \footnote{T-duality exchanges the isometry of one-side with the winding states of the other side. The flavor flux operator becomes to the nontrivial measure factor in mirror path-integral under T-duality, which suggests the right variables one should use in the mirror.}
The constant $C^{g-1}$ is T-duality invariant and will also show up in the mirror Landau-Ginzburg models' correlation functions \cite{Hori:2001ax,Gu:2018fpm}.\\
\section{$SO/O_{+}$ dualities in the mirror}\label{SOOD}
In the mirror, we also integrate out ``matter fields" to the effective theory which depends solely on $\Sigma$ fields.
\subsection{Mirror to $O_+(k)/SO(k)$ gauge theory}
\label{sect:o+k:mirror}
Following \cite{Gu:2018fpm}[sections 9, 10], the mirror is an
orbifold of a Landau-Ginzburg model with fields
\begin{itemize}
\item $Y_{i \alpha}$, $i \in \{1, \cdots, N\}$, $\alpha \in \{1, \cdots, k \}$,
\item $X_{\mu \nu} = \exp\left(-Z_{\mu \nu} \right)$,
$X_{\mu \nu} = X_{\nu \mu}^{-1}$, $\mu \nu \in \{1, \cdots, k\}$
(excluding $X_{2a-1,2a}$),
\item $\Sigma_a$, $a \in \{1, \cdots, N_{c}\}$ where for $k$ even,
$k=2N_{c}$, and for $k$ odd, $k = 2N_{c}+1$,
\end{itemize}
with superpotential
\begin{eqnarray}
W & = &
\sum_{a=1}^{N_{c}} \Sigma_a \left(
\sum_{i \alpha \beta} \rho^a_{i \alpha \beta} Y_{i \beta} \: + \:
\sum_{\mu < \nu; \mu' , \nu'} \alpha^a_{\mu \nu, \mu' \nu'} Z_{\mu' \nu'}
\: - \: t \right)
\nonumber \\
& & \: + \:
\sum_{i \alpha} \exp\left( - Y_{i \alpha} \right)
\: + \:
\sum_{\mu < \nu} X_{\mu \nu}
\nonumber \\
& & \: - \:
\sum_{i, \alpha} m_i  Y_{i \alpha}.
\end{eqnarray}
There is no continuous FI parameter, but we retain $t$ to allow for the possibility of a discrete theta angle. The dualities discussed in \cite{Hori:2011pd} only apply to regular theories in the sense of \cite{Hori:2011pd}, which means we turn on mod 2 theta angle, if $N-k$ is even and we turn off the mod 2 theta angle, if $N-k$ is odd. In the superpotential above,
we take
\begin{eqnarray}
\rho^a_{i \alpha \beta} & = &  \delta_{\alpha, 2a-1}
\delta_{\beta, 2a} - \delta_{\beta, 2a-1} \delta_{\alpha, 2a} ,
\label{eq:so2k1:rho-defn}
\\
\alpha^a_{\mu \nu, \mu' \nu'} & = &
 \delta_{\nu \nu'} \left( \delta_{\mu, 2a-1} \delta_{\mu', 2a} -
\delta_{\mu, 2a} \delta_{\mu', 2a-1} \right)
\nonumber \\
& & \hspace*{0.5in}
\: + \:
 \delta_{\mu \mu'} \left( \delta_{\nu, 2a-1} \delta_{\nu', 2a} -
\delta_{\nu, 2a} \delta_{\nu', 2a-1} \right).
\label{eq:so2k1:alpha-defn}
\end{eqnarray}

Let us split the discussion into four classes:
\begin{itemize}
\item \noindent \textbf{$k=2N_{c}$, $N=2N_{f}$} \quad\quad For a regular theory, we turn on the mod 2 theta angle such that $q=e^{-t}=-1$. From the vacua equation,
\begin{equation}\label{VE}\nonumber
  \exp\left(\frac{\partial W_{eff}(\sigma)}{\partial \sigma_{a}}\right)=1.
\end{equation}
We can obtain
\begin{equation}\label{VE22}
  \prod^{N}_{i=1}\left(\sigma_{a}-m_{i}\right)=(-1)^{k+1}\prod^{N}_{i=1}\left(-\sigma_{a}-m_{i}\right),
\end{equation}
with the excluded locus
\begin{equation}\label{ECL22}
  \sigma_{a}\neq \pm m_{i}, \quad\quad\quad \sigma_{a}\neq\sigma_{b} \quad a\neq b.
\end{equation}
See \cite{Hori:2011pd, Gu:2019zkw}. The degree $N$ polynomial (\ref{VE22}) has $N$ different roots, of the form
\begin{equation}\label{VE22R}
  \sigma=\pm\widehat{\sigma}_{1},\cdots,\pm\widehat{\sigma}_{N_{f}},
\end{equation}
where for generic mass parameters, all of the roots are non-zero. For $O_{+}(k)$, none of these roots intersect with fixed point locus of the orbifold, and for each nonzero root $\widehat{\sigma}_{p}$, the two signs $\pm\widehat{\sigma}_{p}$ are exchanged by a $\mathbb{Z}_{2}$ subgroup of the orbifold group, so define the same vacuum, we count
\begin{equation}\label{}\nonumber
  \left( \begin{array}{c} N_{f} \\ N_{c} \end{array} \right).
\end{equation}
Like \cite{Closset:2017vvl}, we denote those vacua as $\mathcal{S}(N_{c}, N_{f})$. While for $SO(k)$ case, the $\mathbb{Z}_{2}$ is a global symmetry such that we should treat $\pm\widehat{\sigma}_{p}$ are two different vacua, and we count
\begin{equation}\label{}\nonumber
  2\left( \begin{array}{c} N_{f} \\ N_{c} \end{array} \right).
\end{equation}
\\
The computation of the correlation function of $SO(k)$ case:
\begin{equation}\label{CLSOk}
\langle \cO_{0}+\rm{Pf}(\sigma)  \cO_{1}(\sigma^{2}_{a})\rangle_{g}=\sum_{p\in \mathcal{S}(N_{c}, N_{f})}2\cO_{0}(\widehat{\sigma}^{2}_{p}){\textit{H}}(\widehat{\sigma}_{p})^{g-1},
\end{equation}
with
\begin{eqnarray} \label{H22}
   H(\widehat{\sigma}_{a})&=&C\frac{\prod_{i,a}\left(m^{2}_{i}-\widehat{\sigma}^{2}_{a}\right)}{\prod_{a\neq b}\left(\widehat{\sigma}^{2}_{a}-\widehat{\sigma}^{2}_{b}\right)}\left[\prod_{a}\sum_{i}\left(\frac{1}{\widehat{\sigma}_{a}-m_{i}}-\frac{1}{\widehat{\sigma}_{a}+m_{i}}\right)\right] \\\nonumber
  &=& C\frac{\prod_{i,a}\left(m^{2}_{i}-\widehat{\sigma}^{2}_{a}\right)}{\prod_{a\neq b}\left(\widehat{\sigma}^{2}_{a}-\widehat{\sigma}^{2}_{b}\right)}\frac{\prod_{a}P'\left(\widehat{\sigma}_{a}\right)}{\prod_{i,a}\left(\widehat{\sigma}_{a}-m_{i}\right)},
\end{eqnarray}
where we have used the fact in \cite{Closset:2017vvl} [page 23, equation (4.20)], the $P(\sigma)$ is
\begin{equation}\label{P}
  P(\sigma)=2\prod^{N_{f}}_{p=1}\left(\sigma^{2}-\widehat{\sigma}_{p}^{2}\right).
\end{equation}
One can notice that the Pfaffian of $\sigma$, $\rm{Pf}(\sigma)=\prod_{a}\sigma_{a}$, is a physical observable in the $SO(k)$ theory in contrast to the $O_{+}(k)$ group. Because the $\mathbb{Z}_{2}$ is a global symmetry in $SO(k)$ theory while it is a gauge symmetry in the $O_{+}(k)$ case. However like comment in \cite{Closset:2017vvl} [page 33-34], the vacua expectation value of this operator is zero as well.\\

Next, we can write out the correlation function of $O_{+}(k)$, because the vacua do not intersect with the orbifold fixed point, thus we do not need to consider the twisted sectors.
\begin{equation}\label{CLOk}
\langle \cO_{0}+\tau\cdot\cO_{1}(\sigma^{2}_{a})\rangle_{g}=\frac{1}{2}\sum_{p\in \mathcal{S}(N_{c}, N_{f})}2\cO_{0}(\widehat{\sigma}^{2}_{p}){H}(\widehat{\sigma}_{p})^{g-1}=\sum_{p\in \mathcal{S}(N_{c}, N_{f})}\cO_{0}(\widehat{\sigma}^{2}_{p}){H}(\widehat{\sigma}_{p})^{g-1}.
\end{equation}
The twist operator $\tau$ corresponds to the Pfaffian operator in the twisted chiral ring which is defined in \cite{Hori:2011pd,Closset:2017vvl}. We also use the fact that $\tau$ operator's correlation function must vanish.

\item \noindent \textbf{$k=2N_{c}$, $N=2N_{f}+1$} \quad\quad For a regular theory, we turn off the mod 2 theta angle such that $q=e^{-t}=1$. From the vacua equation,
\begin{equation}\label{VE}
  \exp\left(\frac{\partial W_{eff}(\sigma)}{\partial \sigma_{a}}\right)=1.
\end{equation}
We can obtain the vacua equation
\begin{equation}\label{VE21}
  \prod^{N}_{i=1}\left(\sigma_{a}-m_{i}\right)=(-1)^{k}\prod^{N}_{i=1}\left(-\sigma_{a}-m_{i}\right),
\end{equation}
with the excluded locus
\begin{equation}\label{ECL21}
  \sigma_{a}\neq \pm m_{i}, \quad\quad\quad \sigma_{a}\neq\sigma_{b} \quad a\neq b.
\end{equation}
This polynomial (\ref{VE21}) has $N$ different roots, of the form
\begin{equation}\label{VE22R}
  \sigma=0,\pm\widehat{\sigma}_{1},\cdots,\pm\widehat{\sigma}_{N_{f}},
\end{equation}
The vacua solutions have $N_{f}$ pairs of non-zero roots and one root at $\sigma=0$. For $O_{+}(k)$, one root intersects with fixed point locus of the $\mathbb{Z}_{2}$ orbifold, and the twisted sector should be taken into account. The non-zero roots of two signs $\pm\widehat{\sigma}_{p}$ are exchanged by a $\mathbb{Z}_{2}$ subgroup of the orbifold group, so define the same vacuum, we count
\begin{equation}\label{}\nonumber
  \left( \begin{array}{c} N_{f} \\ N_{c} \end{array} \right)+2\left( \begin{array}{c} N_{f} \\ N_{c}-1\end{array} \right).
\end{equation}

For $SO(k)$, the $\mathbb{Z}_{2}$ is a global symmetry which means no twisted sector should be taken into account. When we choose, for example, one $\sigma_{1}=0$, the global symmetry $\mathbb{Z}_{2}$ trivially acts on the ground states. However, if count the solutions of $\sigma_{a}$'s are all non-zero roots, we treat $\pm\widehat{\sigma}_{p}$ are two different vacua, we count the vacua number is
\begin{equation}\label{}\nonumber
  2\left( \begin{array}{c} N_{f} \\ N_{c} \end{array} \right)+\left( \begin{array}{c} N_{f} \\ N_{c}-1\end{array} \right).
\end{equation}
The correlation function of $SO(k)$-case can be computed directly
\begin{equation}\label{CLSOk2}
\langle \cO_{0}+\rm{Pf}(\sigma)  \cO_{1}(\sigma^{2}_{a})\rangle_{g}=\sum_{p\in \mathcal{S}(N_{c}, N_{f})}2\cO_{0}(\widehat{\sigma}^{2}_{p}){\textit{H}}(\widehat{\sigma}_{p})^{g-1}+\sum_{p\in \mathcal{S}(N_{c}-1, N_{f})}\cO_{0}(\widehat{\sigma}^{2}_{p}){\textit{H}}(\widehat{\sigma}_{p},0)^{g-1}.
\end{equation}

The correlation function of the case of $O_{+}(k)$ is more complicated and we need to take into account the twisted sector

\begin{eqnarray}\label{CLOk2}\nonumber
 \langle \cO_{0}+\tau\cdot\cO_{1}(\sigma^{2}_{a})\rangle_{g}  &=& \frac{1}{2}\left(\sum_{p\in \mathcal{S}(N_{c}, N_{f})}2\cO_{0}(\widehat{\sigma}^{2}_{p}){H}(\widehat{\sigma}_{p})^{g-1}+{\cal}{v}^{g}\sum_{p\in \mathcal{S}(N_{c}-1, N_{f})}\cO_{0}(\widehat{\sigma}^{2}_{p}){H}(\widehat{\sigma}_{p},0)^{g-1}\right) \\
   &+&\frac{1}{2}\left[\left({\cal}{v}+3{\cal}{w}\right)^{g}-{\cal}{v}^{g}\right]\left( \sum_{p\in \mathcal{S}(N_{c}-1, N_{f})}\cO_{0}(\widehat{\sigma}^{2}_{p}){H}(\widehat{\sigma}_{p},0)^{g-1}\right)\\\nonumber
   &=&\sum_{p\in \mathcal{S}(N_{c}, N_{f})}\cO_{0}(\widehat{\sigma}^{2}_{p}){H}(\widehat{\sigma}_{p})^{g-1}+\sum_{p\in \mathcal{S}(N_{c}-1, N_{f})}2\cO_{0}(\widehat{\sigma}^{2}_{p})\left({4H}(\widehat{\sigma}_{p},0)\right)^{g-1}
\end{eqnarray}

We have taken into account the twisted sector in the above equation following section \ref{TLGMMO}, and we have set ${\cal}{v}={\cal}{w}=1$ by discussing the genus-zero as well as genus-one correlation functions. The factor $H(\sigma_{p})$ is
\begin{equation}\label{Hf21}
  H(\widehat{\sigma}_{p})= C\frac{\prod_{i,a}\left(m^{2}_{i}-\widehat{\sigma}^{2}_{a}\right)}{\prod_{a\neq b}\left(\widehat{\sigma}^{2}_{a}-\widehat{\sigma}^{2}_{b}\right)}\frac{\prod_{a}Q'\left(\widehat{\sigma}_{a}\right)}{\prod_{i,a}\left(\widehat{\sigma}_{a}-m_{i}\right)},
\end{equation}
 where $Q(\sigma)$ is
 \begin{equation}\label{Q}
  Q(\sigma)=2\sigma\prod^{N_{f}}_{p=1}\left(\sigma^{2}-\widehat{\sigma}_{p}^{2}\right).
 \end{equation}
 The factor $H(\widehat{\sigma}_{p},0)$ is
 \begin{eqnarray}\label{Hf210}
  H(\widehat{\sigma}_{p},0) &=&C \frac{\prod_{i,a}\left(m^{2}_{i}-\widehat{\sigma}^{2}_{a}\right)}{\prod_{a\neq b}\left(\widehat{\sigma}^{2}_{a}-\widehat{\sigma}^{2}_{b}\right)}\left[\prod_{a}\sum_{i}\left(\frac{1}{\widehat{\sigma}_{a}-m_{i}}-\frac{1}{\widehat{\sigma}_{a}+m_{i}}\right)\right]\mid_{\widehat{\sigma}_{N_{c}}=0}  \\\nonumber
   &=&C\frac{\prod_{i,a\neq N_{c}}\left(m^{2}_{i}-\widehat{\sigma}^{2}_{a}\right)\prod_{i}m^{2}_{i}\left(-\sum_{i}\frac{2}{m_{i}}\right)\prod_{a\neq N_{c}}Q'\left(\widehat{\sigma}^{2}_{a}\right)}{\prod_{i,a\neq N_{c}}\left(\widehat{\sigma}_{a}-m_{i}\right)\prod_{a\neq N_{c}}\widehat{\sigma}^{2}_{a}\prod_{b\neq N_{c}}\left(-\widehat{\sigma}^{2}_{b}\right)\prod _{a\neq b\neq N_{c}}\left(\widehat{\sigma}^{2}_{a}-\widehat{\sigma}^{2}_{b}\right)}\\\nonumber
   &=&H(\widehat{\sigma}_{p})\mid_{\widehat{\sigma}_{N_{c}}=0}
\end{eqnarray}

\item \noindent \textbf{$k=2N_{c}+1$, $N=2N_{f}$} \quad\quad For a regular theory, we turn off the mod 2 theta angle such that $q=e^{-t}=1$. From the vacua equation,
\begin{equation}\label{VE1}\nonumber
  \exp\left(\frac{\partial W_{eff}(\sigma)}{\partial \sigma_{a}}\right)=1.
\end{equation}
One can get
\begin{equation}\label{VE12}
  \prod^{N}_{i=1}\left(\sigma_{a}-m_{i}\right)=(-1)^{k}\prod^{N}_{i=1}\left(-\sigma_{a}-m_{i}\right),
\end{equation}
with the excluded locus
\begin{equation}\label{ECL12}
  \sigma_{a}\neq \pm m_{i}, \quad\quad \sigma_{a}\neq\sigma_{b} \quad a\neq b,\quad\quad \sigma_{a}\neq 0
\end{equation}
This degree $N$ polynomial (\ref{VE1}) has $\frac{N}{2}$ pairs non zero roots, of the form
\begin{equation}\label{VE22R}
  \pm\widehat{\sigma}_{1},\cdots,\pm\widehat{\sigma}_{N_{f}}.
\end{equation}
For $O_{+}(k)$, for each nonzero root $\widehat{\sigma}_{p}$, the two signs $\pm\widehat{\sigma}_{p}$ are exchanged by the orbifold group, so define the same vacua. However, due to the even number of the flavor matter fields' last components \cite{Hori:2011pd,Gu:2019zkw}, we count
\begin{equation}\label{}\nonumber
  2\left( \begin{array}{c} N_{f} \\ N_{c} \end{array} \right).
\end{equation}
For $SO(k)$ case, we count
\begin{equation}\label{}\nonumber
  \left( \begin{array}{c} N_{f} \\ N_{c} \end{array} \right).
\end{equation}
\\
For $SO(k)$, the correlation function again can be computed directly which is
\begin{equation}\label{CLSOk3}
\langle \cO_{0}+\rm{Pf}(\sigma)  \cO_{1}(\sigma^{2}_{a})\rangle_{g}=\sum_{p\in \mathcal{S}(N_{c}, N_{f})}\cO_{0}(\widehat{\sigma}^{2}_{p}){\textit{H}}(\widehat{\sigma}_{p})^{g-1}.
\end{equation}
Now we can deduce the correlation function of the case of $O_{+}(k)$. Because the vacua intersect with orbifold fixed point, we need to sum the twisted sectors:
\begin{equation}\label{CLOk3}
\langle \cO_{0}+\tau\cdot\cO_{1}(\sigma^{2}_{a})\rangle_{g}=\frac{\left({\cal}{v}+3{\cal}{w}\right)^{g}}{2}\sum_{p\in \mathcal{S}(N_{c}, N_{f})}\cO_{0}(\widehat{\sigma}^{2}_{p}){H}(\widehat{\sigma}_{p})^{g-1}=2\sum_{p\in \mathcal{S}(N_{c}, N_{f})}\cO_{0}(\widehat{\sigma}^{2}_{p})\left(4H(\widehat{\sigma}_{p})\right)^{g-1},
\end{equation}
with
\begin{equation} \label{H12}
   {H}(\widehat{\sigma}_{a})= C \frac{\prod_{i}(-m_{i})\prod_{i,a}\left(m^{2}_{i}-\widehat{\sigma}^{2}_{a}\right)}{\prod_{a}\widehat{\sigma}^{2}_{a}\prod_{a\neq b}\left(\widehat{\sigma}^{2}_{a}-\widehat{\sigma}^{2}_{b}\right)}\frac{\prod_{a}P'\left(\widehat{\sigma}_{a}\right)}{\prod_{i,a}\left(\widehat{\sigma}_{a}-m_{i}\right)}.
\end{equation}
We have set ${\cal}{v}={\cal}{w}=1$ from the study of the sphere as well as torus correlation functions.\\

\item \noindent \textbf{$k=2N_{c}+1$, $N=2N_{f}+1$} \quad\quad For a regular theory, we turn on the mod 2 theta angle such that $q=e^{-t}=-1$.
One has the vacua equation
\begin{equation}\label{VE11}
  \prod^{N}_{i=1}\left(\sigma_{a}-m_{i}\right)=(-1)^{k+1}\prod^{N}_{i=1}\left(-\sigma_{a}-m_{i}\right),
\end{equation}
with the excluded locus
\begin{equation}\label{ECL11}
  \sigma_{a}\neq \pm m_{i}, \quad\quad \sigma_{a}\neq\sigma_{b} \quad\text{for}\quad a\neq b,\quad\quad \sigma_{a}\neq 0.
\end{equation}
This degree $N$ polynomial (\ref{VE12}) has $\frac{(N-1)}{2}$ pairs non zero roots and one root at $\sigma=0$, of the form
\begin{equation}\label{VE22R}
  \sigma=0,\pm\widehat{\sigma}_{1},\cdots,\pm\widehat{\sigma}_{N_{f}}.
\end{equation}
For $O_{+}(k)$, for each nonzero root $\widehat{\sigma}_{p}$, the two signs $\pm\widehat{\sigma}_{p}$ are exchanged by the orbifold group, so define the same vacuum. The zero root locates at the excluded locus that we should avoid. For odd number of the matter field, we count
\begin{equation}\label{}\nonumber
  \left( \begin{array}{c} N_{f} \\ N_{c} \end{array} \right).
\end{equation}
For $SO(k)$ case, we have
\begin{equation}\label{}\nonumber
  \left( \begin{array}{c} N_{f} \\ N_{c} \end{array} \right).
\end{equation}
\\
The correlation function of $SO(k)$ is
\begin{equation}\label{CLSOk4}
\langle \cO_{0}+\rm{Pf}(\sigma)  \cO_{1}(\sigma^{2}_{a})\rangle_{g}=\sum_{p\in \mathcal{S}(N_{c}, N_{f})}\cO_{0}(\widehat{\sigma}^{2}_{p}){\textit{H}}(\widehat{\sigma}_{p})^{g-1}.
\end{equation}
Now we can deduce the correlation function of the case of $O_{+}(k)$, which is
\begin{eqnarray}
  \langle \cO_{0}+\tau\cdot\cO_{1}(\sigma^{2}_{a})\rangle_{g} &=&  \frac{\left({\cal}{v}+3{\cal}{w}\right)^{g}}{2}\sum_{p\in \mathcal{S}(N_{c}, N_{f})}\cO_{0}(\widehat{\sigma}^{2}_{p})H(\widehat{\sigma}_{p})^{g-1}\\\nonumber
   &=& \sum_{p\in \mathcal{S}(N_{c}, N_{f})}\cO_{0}(\widehat{\sigma}^{2}_{p})\left(2H(\widehat{\sigma}_{p})\right)^{g-1},
\end{eqnarray}
with
\begin{equation} \label{H11}
   H(\widehat{\sigma}_{a})=C \frac{\prod_{i}(-m_{i})\prod_{i,a}\left(m^{2}_{i}-\widehat{\sigma}^{2}_{a}\right)}{\prod_{a}\widehat{\sigma}^{2}_{a}\prod_{a\neq b}\left(\widehat{\sigma}^{2}_{a}-\widehat{\sigma}^{2}_{b}\right)}\frac{\prod_{a}Q'\left(\widehat{\sigma}_{a}\right)}{\prod_{i,a}\left(\widehat{\sigma}_{a}-m_{i}\right)}.
\end{equation}
We have set ${\cal}{v}=-1$ and ${\cal}{w}=1$.
\end{itemize}
\subsection{Mirror to dual $SO(N-k+1)/O_+(N-k+1)$ }
The mirror superpotential of these gauge theories is
\begin{eqnarray}
W & = & \sum_{a^{D}=1}^{N^{D}_{c}} \Sigma_a^{D} \left( - \sum_{i, \alpha, \beta}
\rho^{a^{D}}_{i \alpha \beta} \ln \left( W^{i \beta} \right)^2 \: - \:
\sum_{\mu < \nu; \mu', \nu'} \alpha^{a^{D}}_{\mu \nu; \mu' \nu'}
\ln X_{\mu \nu} \: - \: \tilde{t} \right)
\nonumber \\
& &
\: + \: \sum_{i \alpha}  \left( W^{i \alpha} \right)^2
\: + \: \sum_{\mu < \nu} X_{\mu \nu}
\: + \: \sum_{i \leq j} \exp\left( - T_{ij} \right)
\nonumber \\
& &
\: - \: \sum_{i \alpha} 2 m_{i} \ln W^{i \alpha}
\: - \: \sum_{i \leq j} \left( m_i + m_j \right) T_{ij},
\end{eqnarray}
where
\begin{eqnarray}
\rho^{a^{D}}_{i \alpha \beta} & = &  \delta_{\alpha, 2a^{D}-1}
\delta_{\beta, 2a^{D}} - \delta_{\beta, 2a^{D}-1} \delta_{\alpha, 2a^{D}} ,
\label{eq:so2k:rho-defn}
\\
\alpha^{a^{D}}_{\mu \nu, \mu' \nu'} & = &
 \delta_{\nu \nu'} \left( \delta_{\mu, 2a^{D}-1} \delta_{\mu', 2a^{D}} -
\delta_{\mu, 2a^{D}} \delta_{\mu', 2a^{D}-1} \right)
\nonumber \\
& & \hspace*{0.5in}
\: + \:
 \delta_{\mu \mu'} \left( \delta_{\nu, 2a^{D}-1} \delta_{\nu', 2a^{D}} -
\delta_{\nu, 2a^{D}} \delta_{\nu', 2a^{D}-1} \right).
\label{eq:so2k:alpha-defn}
\end{eqnarray}
The case $N-k+1$ odd is described in a formally identical fashion. Here, we define $N^{D}_{c}$ by $N-k+1 = 2N^{D}_{c}+1$, and in the description of the Weyl group, the kernel $K$ is taken to be all of $({\mathbb Z}_2)^{N^{D}_{c}}$.
Its action on the fields is identical to the case $N-k+1$ even -- meaning, for example, that $W^{i, N-k+1} = W^{i,2N^{D}_{c}+1}$ is invariant under the Weyl group orbifold.\\

It is straightforward to verify that on the critical locus, one has the Coulomb branch relation
\begin{equation}   \label{eq:sonk1:mirror:qc}
\prod_{i=1}^N \left( \sigma_a^{D} \: - \: m_i \right)
\: = \:
(-)^{N-k+1} \tilde{q}
\prod_{i=1}^N \left( - \sigma_a^{D} \: - \: m_i \right).
\end{equation}
The excluded locus computed from the critical locus relations for
the $W$ fields above is then
\begin{equation}
\sigma_a^{D} \: \neq \: \pm m_i.
\end{equation}
From the $X$ fields, for $a^{D} \neq b^{D}$,
\begin{equation}   \label{eq:sonk1:mirror:excluded2}
\sigma_a^{D} \: \neq \: \pm \sigma_b^{D},
\end{equation}
and for $N-k+1$ odd, we also have
\begin{equation}    \label{eq:sonk1:mirror:excluded3}
\sigma_a^{D} \: \neq \: 0.
\end{equation}
For dual $SO$ case, one can argue that the orbifold group will exchange roots, but the critical locus
will not intersect the fixed point locus of the orbifold in this mirror, so in this particular mirror we do not have to consider any twisted sector ground states. For dual $O_{+}$, we could have vacua intersect with $\mathbb{Z}_{2}$ orbifold fixed point such that we have taken into account the twisted sector.\\

The computation of correlation functions in the dual theories is almost the same as we did in section \ref{sect:o+k:mirror}, thus we will just list the results. \\
\begin{itemize}
\item \noindent \textbf{$N-k+1=2N^{D}_{c}$, $N=2N_{f}$} \quad\quad For a regular theory, we turn on the mod 2 theta angle. The vacua equation is
\begin{equation}\label{DVE22}
  \prod^{N}_{i=1}\left(\sigma_{a^{D}}-m_{i}\right)=(-1)^{N-k}\prod^{N}_{i=1}\left(-\sigma_{a^{D}}-m_{i}\right),
\end{equation}
with the excluded locus
\begin{equation}\label{DECL22}
  \sigma_{a^{D}}\neq \pm m_{i}, \quad\quad \sigma_{a^{D}}\neq\sigma_{b^{D}} \quad {\rm if} a^{D}\neq b^{D}.
\end{equation}
The degree $N$ polynomial (\ref{DVE22}) has $N$ different roots, of the form
\begin{equation}\label{DVE22R}
  \sigma=\pm\widehat{\sigma}_{1},\cdots,\pm\widehat{\sigma}_{N_{f}},
\end{equation}
where for generic mass parameters, all of the roots are non-zero. For $O_{+}(N-k+1)$, none of these roots intersect with fixed point locus of the orbifold, and for each nonzero root $\widehat{\sigma}_{p^{D}}$, the two signs $\pm\widehat{\sigma}_{p^{D}}$ are exchanged by a $\mathbb{Z}_{2}$ subgroup of the orbifold group, so define the same vacuum, we count
\begin{equation}\label{}\nonumber
  \left( \begin{array}{c} N_{f} \\ N^{D}_{c} \end{array} \right).
\end{equation}
Those vacua can be denoted as $\mathcal{S}(N^{D}_{c}, N_{f})$. While for $SO(N-k+1)$ case, we count
\begin{equation}\label{}\nonumber
  2\left( \begin{array}{c} N_{f} \\ N^{D}_{c} \end{array} \right).
\end{equation}
\\
The correlation function of $SO(N-k+1)$ case is:
\begin{equation}\label{DCLSOk}
\langle \cO_{0}+\rm{Pf}(\sigma)  \cO_{1}(\sigma^{2}_{a^{D}})\rangle_{g}=\sum_{\textit{p}^{D} \in \mathcal{S}(N^{D}_{c}, N_{f})}2\cO_{0}(\widehat{\sigma}^{2}_{ \textit{p}^{D}})\textit{H}_{\textit{D}}(\widehat{\sigma}_{\textit{p}^{D}})^{g-1},
\end{equation}
with
\begin{eqnarray} \label{DH22}
   H_{D}(\widehat{\sigma}_{a^{D}})&=&C_{D}\frac{\prod_{i\leq j}\left(-m_{i}-m_{j}\right)}{\prod_{a^{D}\neq b^{D}}\left(\widehat{\sigma}^{2}_{a^{D}}-\widehat{\sigma}^{2}_{b^{D}}\right)}\left[\prod_{a^{D}}\sum_{i}\left(\frac{1}{\widehat{\sigma}_{a^{D}}+m_{i}}-\frac{1}{\widehat{\sigma}_{a^{D}}-m_{i}}\right)\right] \\\nonumber
  &=& C_{D}\frac{\prod_{i\leq j}\left(-m_{i}-m_{j}\right)}{\prod_{a^{D}\neq b^{D}}\left(\widehat{\sigma}^{2}_{a^{D}}-\widehat{\sigma}^{2}_{b^{D}}\right)}\frac{\prod_{a^{D}}P'\left(\widehat{\sigma}_{a^{D}}\right)}{\prod_{i,a^{D}}\left(\widehat{\sigma}_{a^{D}}+m_{i}\right)}.
\end{eqnarray}
The correlation function for $O_{+}(N-k+1)$ is
\begin{equation}\label{DCLOk}
\langle \cO_{0}+\tau\cdot\cO_{1}(\sigma^{2}_{a^{D}})\rangle_{g}=\sum_{p^{D}\in \mathcal{S}(N^{D}_{c}, N_{f})}\cO_{0}(\widehat{\sigma}^{2}_{p^{D}})H_{D}(\widehat{\sigma}_{p^{D}})^{g-1}.
\end{equation}

\item \noindent \textbf{$N-k+1=2N^{D}_{c}$, $N=2N_{f}+1$} \quad\quad For a regular theory, we turn off the mod 2 theta angle. The vacua equation is
\begin{equation}\label{DVE21}
  \prod^{N}_{i=1}\left(\sigma_{a^{D}}-m_{i}\right)=(-1)^{N-k+1}\prod^{N}_{i=1}\left(-\sigma_{a^{D}}-m_{i}\right),
\end{equation}
with the excluded locus
\begin{equation}\label{DECL21}
  \sigma_{a^{D}}\neq \pm m_{i}, \quad\quad \sigma_{a^{D}}\neq\sigma_{b^{D}} \quad {\rm if} a^{D}\neq b^{D}.
\end{equation}
This polynomial (\ref{DVE21}) has $N$ different roots, of the form
\begin{equation}\label{DVE22R}
  \sigma=0,\pm\widehat{\sigma}_{1},\cdots,\pm\widehat{\sigma}_{N_{f}}.
\end{equation}
For $O_{+}(N-k+1)$, we count the vacua
\begin{equation}\label{}\nonumber
  \left( \begin{array}{c} N_{f} \\ N^{D}_{c} \end{array} \right)+2\left( \begin{array}{c} N_{f} \\ N^{D}_{c}-1\end{array} \right).
\end{equation}

For $SO(N-k+1)$, the vacua number is
\begin{equation}\label{}\nonumber
  2\left( \begin{array}{c} N_{f} \\ N^{D}_{c} \end{array} \right)+\left( \begin{array}{c} N_{f} \\ N^{D}_{c}-1\end{array} \right).
\end{equation}
The correlation function for the $SO(N-k+1)$-case is
\begin{equation}\label{DCLSOk2}
\langle \cO_{0}+\rm{Pf}(\sigma)  \cO_{1}(\sigma^{2}_{a^{D}})\rangle_{g}=\sum_{\textit{p}^{D}\in \mathcal{S}(N^{D}_{c}, N_{f})}2\cO_{0}(\widehat{\sigma}^{2}_{\textit{p}^{D}})\textit{H}_{\textit{D}}(\widehat{\sigma}_{\textit{p}^{D}})^{g-1}+\sum_{\textit{p}^{D}\in \mathcal{S}(N^{D}_{c}-1, N_{f})}\cO_{0}(\widehat{\sigma}^{2}_{\textit{p}^{D}})\textit{H}_{\textit{D}}(\widehat{\sigma}_{\textit{p}^{D}},0)^{g-1}.
\end{equation}
The factor $H_{D}$ is
\begin{equation}\label{DHf21}
  H_{D}(\widehat{\sigma}_{p^{D}})= C_{D}\frac{\prod_{i\leq j}\left(-m_{i}-m_{j}\right)}{\prod_{a^{D}\neq b^{D}}\left(\widehat{\sigma}^{2}_{a^{D}}-\widehat{\sigma}^{2}_{b^{D}}\right)}\frac{\prod_{a}Q'\left(\widehat{\sigma}_{a^{D}}\right)}{\prod_{i,a^{D}}\left(\widehat{\sigma}_{a^{D}}+m_{i}\right)},
\end{equation}
 where $Q(\sigma)$ is
 \begin{equation}\label{DQ}
  Q(\sigma)=2\sigma\prod^{N_{f}}_{p^{D}=1}\left(\sigma^{2}-\widehat{\sigma}_{p^{D}}^{2}\right).
 \end{equation}\\

The correlation function for the $O_{+}(N-k+1)$-case is
\begin{equation}\label{DCLOk2}
\langle \cO_{0}+\tau\cdot\cO_{1}(\sigma^{2}_{a^{D}})\rangle_{g}=\sum_{p^{D}\in \mathcal{S}(N^{D}_{c}, N_{f})}\cO_{0}(\widehat{\sigma}^{2}_{p^{D}}){H_{D}}(\widehat{\sigma}_{p^{D}})^{g-1}+\sum_{p^{D}\in \mathcal{S}(N^{D}_{c}-1, N_{f})}2\cO_{0}(\widehat{\sigma}^{2}_{p^{D}})\left(4H_{D}(\widehat{\sigma}_{p^{D}},0)\right)^{g-1}.
\end{equation}

\item \noindent \textbf{$N-k+1=2{N}^{D}_{c}+1$, $N=2N_{f}$} \quad\quad For a regular theory, we turn off the mod 2 theta angle. The vacua equation is
\begin{equation}\label{DVE12}
  \prod^{N}_{i=1}\left(\sigma_{a^{D}}-m_{i}\right)=(-1)^{N-k}\prod^{N}_{i=1}\left(-\sigma_{a^{D}}-m_{i}\right),
\end{equation}
with the excluded locus
\begin{equation}\label{DECL12}
  \sigma_{a^{D}}\neq \pm m_{i}, \quad\quad \sigma_{a^{D}}\neq\sigma_{b^{D}} \quad {\rm if} a^{D}\neq b^{D},\quad\quad \sigma_{a^{D}}\neq 0
\end{equation}
For $O_{+}(N-k+1)$, we count the vacua number
\begin{equation}\label{}\nonumber
  2\left( \begin{array}{c} N_{f} \\ N^{D}_{c} \end{array} \right).
\end{equation}
While for $SO(N-k+1)$ case, we count
\begin{equation}\label{}\nonumber
  \left( \begin{array}{c} N_{f} \\ N^{D}_{c} \end{array} \right).
\end{equation}
\\
The correlation function of $SO(N-k+1)$ is
\begin{equation}\label{DCLSOk3}
\langle \cO_{0}+\rm{Pf}(\widetilde{\sigma})  \cO_{1}(\sigma^{2}_{a^{D}})\rangle_{g}=\sum_{\textit{p}^{D}\in \mathcal{S}(N^{D}_{c}, N_{f})}\cO_{0}(\widehat{\sigma}^{2}_{\textit{p}^{D}}){\textit{H}_{\textit{D}}}(\widehat{\sigma}_{\textit{p}^{D}})^{g-1}.
\end{equation}
with
\begin{equation} \label{DH12}
   H_{D}(\widehat{\sigma}_{a^{D}})= C_{D}\frac{\prod_{i\leq j}\left(-m_{i}-m_{j}\right)}{\prod_{a^{D}}\widehat{\sigma}^{2}_{a^{D}}\prod_{a^{D}\neq b^{D}}\left(\widehat{\sigma}^{2}_{a^{D}}-\widehat{\sigma}^{2}_{b^{D}}\right)}\frac{\prod_{a^{D}}P'\left(\widehat{\sigma}_{a^{D}}\right)}{\prod_{i,a^{D}}\left(\widehat{\sigma}_{a^{D}}+m_{i}\right)}.
\end{equation}
\\
The correlation function of the case of $O_{+}(N-k+1)$ is
\begin{equation}\label{DCLOk3}
\langle \cO_{0}+\tau\cdot\cO_{1}(\sigma^{2}_{a^{D}})\rangle_{g}=\sum_{p\in \mathcal{S}(N^{D}_{c}, N_{f})}2\cO_{0}(\widehat{\sigma}^{2}_{p})\left(4H_{D}(\widehat{\sigma}_{p})\right)^{g-1},
\end{equation}

\item \noindent \textbf{$N-k+1=2N^{D}_{c}+1$, $N=2N_{f}+1$} \quad\quad For a regular theory, we turn on the mod 2 theta angle.
One has the vacua equation
\begin{equation}\label{DVE11}
  \prod^{N}_{i=1}\left(\sigma_{a^{D}}-m_{i}\right)=(-1)^{N-k}\prod^{N}_{i=1}\left(-\sigma_{a^{D}}-m_{i}\right),
\end{equation}
with the excluded locus
\begin{equation}\label{DECL11}
  \sigma_{a^{D}}\neq \pm m_{i}, \quad\quad \sigma_{a^{D}}\neq\sigma_{b^{D}} \quad {\rm if} a^{D}\neq b^{D},\quad\quad \sigma_{a^{D}}\neq 0
\end{equation}
For $O_{+}(N-k+1)$, we count the vacua
\begin{equation}\label{}\nonumber
  \left( \begin{array}{c} N_{f} \\ N^{D}_{c} \end{array} \right).
\end{equation}
While for $SO(N-k+1)$ case, we count
\begin{equation}\label{}\nonumber
  \left( \begin{array}{c} N_{f} \\ N^{D}_{c} \end{array} \right).
\end{equation}
\\
For $SO(N-k+1)$, the correlation function is
\begin{equation}\label{DCLSOk4}
\langle \cO_{0}+\rm{Pf}(\sigma)  \cO_{1}(\sigma^{2}_{a^{D}})\rangle_{g}=\sum_{\textit{p}^{D}\in \mathcal{S}(N^{D}_{c}, N_{f})}\cO_{0}(\widehat{\sigma}^{2}_{\textit{p}^{D}})H_{D}(\widehat{\sigma}_{\textit{p}^{D}})^{g-1},
\end{equation}
with
\begin{equation} \label{DH11}
   H_{D}(\widehat{\sigma}_{a^{D}})=C_{D} \frac{\prod_{i\leq j}\left(-m_{i}-m_{j}\right)}{\prod_{a^{D}}\widehat{\sigma}^{2}_{a^{D}}\prod_{a^{D}\neq b^{D}}\left(\widehat{\sigma}^{2}_{a^{D}}-\widehat{\sigma}^{2}_{b^{D}}\right)}\frac{\prod_{a^{D}}Q'\left(\widehat{\sigma}_{a^{D}}\right)}{\prod_{i,a^{D}}\left(\widehat{\sigma}_{a^{D}}+m_{i}\right)}.
\end{equation}\\

The correlation function for $O_{+}(N-k+1)$ is
\begin{equation}\label{DCLOk4}
\langle \cO_{0}+\tau\cdot\cO_{1}(\sigma^{2}_{a^{D}})\rangle_{g}=\sum_{\textit{p}^{D}\in \mathcal{S}(N^{D}_{c}, N_{f})}\cO_{0}(\widehat{\sigma}^{2}_{\textit{p}^{D}})\left(2H_{D}(\widehat{\sigma}_{\textit{p}^{D}})\right)^{g-1}.
\end{equation}
\end{itemize}
\subsection{Comparison of the correlation functions of dual theories}\label{CCFD}
We only do the comparison of the dual pairs $SO(k)\leftrightarrow O_{+}(N-k+1)$ in this section. The match of dual pairs $O_{+}(k)\leftrightarrow SO(N-k+1)$ is the same as the previous dual pairs which we leave this to the interested reader.\\
\begin{itemize}
\item \noindent $SO(2N_{c})\leftrightarrow O_{+}(2N^{D}_{c})$, $N=2N_{f}+1$, $N^{D}_{c}=N_{f}-N_{c}+1$:\\
Map of vacua: the discussion here is the same as in \cite{Closset:2017vvl}. It is more clear to map the vacua if we focus on the polynomial $Q(z)$, which is
\begin{equation}\label{}\nonumber
  Q(z)=2z\prod^{N_{f}}_{p=1}\left(z^{2}-\widehat{z}^{2}_{p}\right),
\end{equation}
where $z$ stands for $\sigma$ in the $SO(2N_{c})$ theory, while is $\sigma_{D}$ in the dual theory.

There are three types of ground states in the $SO(2N_{c})$ theory, two types of them consist of a nontrivial representation of $\mathbb{Z}_{2}$ global symmetry of $SO(2N_{c})$ theory. made the identification by the Weyl orbifold, we can find two types of vacua represented by $N_{c}$-tuples:
\begin{eqnarray}\nonumber
   \left(\widehat{\sigma}_{p_{1}},\cdots,\widehat{\sigma}_{p_{N_{c}-1}},\pm\widehat{\sigma}_{p_{N_{c}}} \right), \quad\quad\quad\quad&&  p_{1}<\cdots<p_{N_{c}} \quad\quad p_{a}\in[N_{f}]
\end{eqnarray}
These vacua are associated to each element of $\mathcal{S}(N_{c}, N_{f})$ . The third type of the vacua is represented by $\left(N_{c}-1\right)$-tuple, which is
\begin{equation}\label{}\nonumber
  \left(\widehat{\sigma}_{p_{1}},\cdots,\widehat{\sigma}_{p_{N_{c}-1}}, 0 \right), \quad\quad  p_{1}<\cdots<p_{N_{c}-1} \quad\quad p_{a}\in[N_{f}].
\end{equation}
These vacua can be denoted as $\mathcal{S}(N_{c}-1, N_{f})$.\\

The dual theory, there are two types of vacua. One of them is intersecting with $\mathbb{Z}_{2}$ orbifold fixed point and we should take into account the twisted sector, these ground states can be expressed in terms of a length $N^{D}_{c}-1$ vector $p^{D}\in\mathcal{S}(N^{D}_{c}-1, N_{f})$. The representative tuples of roots are
\begin{equation}\label{}\nonumber
  \left(\widehat{\sigma}_{p^{D}_{1}},\cdots,\widehat{\sigma}_{p^{D}_{N^{D}_{c}-1}}, 0 \right), \quad\quad  p^{D}_{1}<\cdots<p^{D}_{N^{D}_{c}-1} \quad\quad p^{D}_{a}\in[N_{f}].
\end{equation}
The other type of ground states can be expressed in terms of $N^{D}_{c}$ vector $\alpha^{D}\in\mathcal{S}(N^{D}_{c}, N_{f})$:
\begin{equation}\label{}\nonumber
  \left(\widehat{\sigma}_{p^{D}_{1}},\cdots,\widehat{\sigma}_{p^{D}_{N^{D}_{c}}} \right), \quad\quad\quad\quad p^{D}_{1}<\cdots<p^{D}_{N^{D}_{c}} \quad\quad p^{D}_{a}\in[N_{f}].
\end{equation}
Follow \cite{Hori:2011pd,Closset:2017vvl}, the dual map is
\begin{eqnarray}\nonumber
  \mathcal{S}(N_{c}, N_{f}) &\Leftrightarrow& \mathcal{S}(N^{D}_{c}-1, N_{f}) \\\nonumber
   \mathcal{S}(N_{c}-1, N_{f}) &\Leftrightarrow&  \mathcal{S}(N^{D}_{c}, N_{f}).
\end{eqnarray}
\\
\noindent\textbf{Comparison of the B-twisted correlation functions}:
Recall the correlation function for the $SO(2N_{c})$ with odd number matter fields is equation (\ref{CLSOk2}),
\begin{equation}\label{} \nonumber
\langle \cO_{0}+\rm{Pf}(\sigma)  \cO_{1}(\sigma^{2}_{a})\rangle_{g}=\sum_{p\in \mathcal{S}(N_{c}, N_{f})}2\cO_{0}(\widehat{\sigma}^{2}_{p})\textit{H}(\widehat{\sigma}_{p})^{g-1}+\sum_{p\in \mathcal{S}(N_{c}-1, N_{f})}\cO_{0}(\widehat{\sigma}^{2}_{p})\textit{H}(\widehat{\sigma}_{p}, 0)^{g-1}.
\end{equation}
While the correlation function of the dual $O_{+}(2N^{D}_{c})$ case is given by equation (\ref{DCLOk2})
\begin{equation}\label{}\nonumber
\langle \cO_{0}+\tau\cdot\cO_{1}(\sigma^{2}_{a})\rangle_{g}=\sum_{p\in \mathcal{S}(N^{D}_{c}, N_{f})}\cO_{0}(\widehat{\sigma}^{2}_{p})H_{D}(\widehat{\sigma}_{p})^{g-1}+\sum_{p\in \mathcal{S}(N^{D}_{c}-1, N_{f})}2\cO_{0}(\widehat{\sigma}^{2}_{p})\left(4H_{D}(\widehat{\sigma}_{p},0)\right)^{g-1}.
\end{equation}
It is obvious that the match of the correlation functions can be transferred to the match of Hessian factors, and indeed follow the appendix A of \cite{Closset:2017vvl}\footnote{One can prove the algebraic identities of that appendix by using the Vieta's formulas to the equation (A.1)}, one can prove that
\begin{equation}\label{H21m}
  \frac{H(\widehat{\sigma}_{p_{a}})}{4H_{D}(\widehat{\sigma}_{p^{D}_{a^{D}}},0)}=\frac{H(\widehat{\sigma}_{p_{a}},0)}{H_{D}(\widehat{\sigma}_{p^{D}_{a^{D}}})}=\frac{C}{C_{D}}2^{-4N^{D}_{c}}e^{i\pi \left(N_{c}N^{D}_{c}+\nu\right)}=1,
\end{equation}
if we impose the following relation between the contact terms
\begin{equation}\label{}\nonumber
 C_{D}=2^{-4N^{D}_{c}}e^{i\pi \left(N_{c}N^{D}_{c}+\nu\right)}C.
\end{equation}
The integer $\nu$ can be uniquely determined by the mass parameters.
\\

\item \noindent $SO(2N_{c})\leftrightarrow O_{+}(2N^{D}_{c}+1)$, $N=2N_{f}$, $N^{D}_{c}=N_{f}-N_{c}+1$:\\
Map of vacua: the polynomial determines the map from vacua of one side to vacua of the other side explicitly
\begin{equation}\label{}\nonumber
  P(z)=2\prod^{N_{f}}_{p=1}\left(z^{2}-\widehat{z}^{2}_{p}\right).
\end{equation}
For the case of $SO(2N_{c})$, the vacua can be represented by $N_{c}$-tuples of roots of $P(\sigma)$ that satisfy the excluded locus condition:
\begin{eqnarray}\nonumber
   \left(\widehat{\sigma}_{p_{1}},\cdots,\pm\widehat{\sigma}_{p_{N_{c}}}\right)&& \quad\quad p_{1}<\cdots<p_{N_{c}}\quad\quad\quad p_{a}\in [N_{f}]
\end{eqnarray}
These vacua are consisting of the nontrivial representation of the global $\mathbb{Z}_{2}$ symmetry, which we denote as $\mathcal{S}(N_{c}, N_{f})$. \\

In the dual theory, we only have one type of vacua which can be represented by
\begin{equation}\label{}\nonumber
  \left(\widehat{\sigma}_{p^{D}_{1}},\cdots,\widehat{\sigma}_{p^{D}_{N^{D}_{c}}}\right), \quad\quad \widehat{\sigma}_{1}<\cdots<\widehat{\sigma}_{N^{D}_{c}}, \quad\quad p^{D}_{\overline{a}}\in [N_{f}].
\end{equation}
The vacua can be denoted by ${\cal} S(N^{D}_{c}, N_{f})$. The theory has a $\mathbb{Z}_{2}$ orbifold such that we should take into account the twisted sector. One can find the vacua match obviously. \\

The correlation function of the case of $SO(2N_{c})$ is \ref{CLSOk}:
\begin{equation}\nonumber
\langle \cO_{0}+\rm{Pf}(\sigma)  \cO_{1}(\sigma^{2}_{a})\rangle_{g}=\sum_{\textit{p}\in \mathcal{S}(N_{c}, N_{f})}2\cO_{0}(\widehat{\sigma}^{2}_{\textit{p}})\textit{H}(\widehat{\sigma}_{\textit{p}})^{g-1},
\end{equation}
  with
  \begin{eqnarray} \label{H22}\nonumber
   H(\widehat{\sigma}_{a})&=&C\frac{\prod_{i,a}\left(m^{2}_{i}-\widehat{\sigma}^{2}_{a}\right)}{\prod_{a\neq b}\left(\widehat{\sigma}^{2}_{a}-\widehat{\sigma}^{2}_{b}\right)}\left[\prod_{a}\sum_{i}\left(\frac{1}{\widehat{\sigma}_{a}-m_{i}}-\frac{1}{\widehat{\sigma}_{a}+m_{i}}\right)\right] \\\nonumber
  &=& C\frac{\prod_{i,a}\left(m^{2}_{i}-\widehat{\sigma}^{2}_{a}\right)}{\prod_{a\neq b}\left(\widehat{\sigma}^{2}_{a}-\widehat{\sigma}^{2}_{b}\right)}\frac{\prod_{a}P'\left(\widehat{\sigma}_{a}\right)}{\prod_{i,a}\left(\widehat{\sigma}_{a}-m_{i}\right)}.
\end{eqnarray}

While the correlation function of the case of $O_{+}(2N^{D}_{c}+1)$ is equation (\ref{DCLOk3})
\begin{equation}\nonumber
\langle \cO_{0}+\tau\cdot\cO_{1}(\sigma^{2}_{a})\rangle_{g}=\sum_{p^{D}\in \mathcal{S}(N^{D}_{c}, N_{f})}2\cO_{0}(\widehat{\sigma}^{2}_{p^{D}})\left(4H_{D}(\widehat{\sigma}_{p^{D}})\right)^{g-1},
\end{equation}
with
\begin{equation} \label{DH12}\nonumber
  H_{D}(\widehat{\sigma}_{a^{D}})= C_{D}\frac{\prod_{i\leq j}\left(-m_{i}-m_{j}\right)}{\prod_{a^{D}}\widehat{\sigma}^{2}_{a^{D}}\prod_{a^{D}\neq b^{D}}\left(\widehat{\sigma}^{2}_{a^{D}}-\widehat{\sigma}^{2}_{b^{D}}\right)}\frac{\prod_{a^{D}}P'\left(\widehat{\sigma}_{a^{D}}\right)}{\prod_{i,a^{D}}\left(\widehat{\sigma}_{a^{D}}+m_{i}\right)}.
\end{equation}
The match of the correlation function can be transferred into the match of the Hessian factor, and one can easily find that
\begin{equation}\label{}\nonumber
  \frac{H(\widehat{\sigma}_{p})}{4H_{D}(\widehat{\sigma}_{p^{D}})}=\frac{C}{C_{D}}2^{-4N^{D}_{c}-2}e^{i\pi \left(N_{c}N^{D}_{c}+\nu\right)}=1,
\end{equation}
if
\begin{equation}\label{}\nonumber
 C_{D}=2^{-4N^{D}_{c}-2}e^{i\pi \left(N_{c}N^{D}_{c}+\nu\right)}C.
\end{equation}
\\

\item \noindent $SO(2N_{c}+1)\leftrightarrow O_{+}(2N^{D}_{c})$, $N=2N_{f}$, $N^{D}_{c}=N_{f}-N_{c}$:\\
Map of vacua: the $P(z)$ is
\begin{equation}\label{}\nonumber
  P(z)=2\prod^{N_{f}}_{p=1}\left(z^{2}-\widehat{z}^{2}_{p}\right).
\end{equation}
The ground state is represented by a tuple of roots:

\begin{equation}\label{}\nonumber
  \left(\widehat{\sigma}_{p_{1}},\cdots,\widehat{\sigma}_{p_{N_{c}}}\right),\quad\quad p_{a}\in {\cal} S(N_{c}, N_{f}).
\end{equation}
In the dual theory, each ground state can be represented by a tuple of roots:
\begin{equation}\label{}\nonumber
   \left(\widehat{\sigma}_{p^{D}_{1}},\cdots,\widehat{\sigma}_{p^{D}_{N_{c}}}\right),\quad\quad p^{D}_{a}\in {\cal} S(N^{D}_{c}, N_{f}).
\end{equation}
So the vacua of the two dual theories match.\\
The correlation function for $SO(2N_{c}+1)$ is
\begin{equation}\label{CLSOk3}\nonumber
\langle \cO_{0}+\rm{Pf}(\sigma)  \cO_{1}(\sigma^{2}_{a})\rangle_{g}=\sum_{p\in \mathcal{S}(N_{c}, N_{f})}\cO_{0}(\widehat{\sigma}^{2}_{p})\textit{H}(\widehat{\sigma}_{p})^{g-1}.
\end{equation}
The correlation function for the $O_{+}(2N^{D}_{c})$ case is
\begin{equation}\label{DCLOk3}\nonumber
\langle \cO_{0}+\tau\cdot\cO_{1}(\sigma^{2}_{a})\rangle_{g}=\sum_{p^{D}\in \mathcal{S}(N^{D}_{c}, N_{f})}\left(\cO_{0}(\widehat{\sigma}^{2}_{p^{D}})\right)H(\widehat{\sigma}_{p^{D}})^{g-1},
\end{equation}
One can find the match of the Hessian
\begin{equation}\label{}\nonumber
  \frac{H(\widehat{\sigma}_{p})}{H(\widehat{\sigma}_{p^{D}})}=\frac{C}{C_{D}}2^{-4N^{D}_{c}}e^{i\pi \left(N_{c}N^{D}_{c}+\nu+N_{f}\right)}=1,
\end{equation}
where we have required
\begin{equation}\label{}\nonumber
  C_{D}=2^{-4N^{D}_{c}}e^{i\pi \left(N_{c}N^{D}_{c}+\nu+N_{f}\right)}C.
\end{equation}
\\
 \item \noindent $SO(2N_{c}+1)\leftrightarrow O_{+}(2N^{D}_{c}+1)$, $N=2N_{f}+1$, $N^{D}_{c}=N_{f}-N_{c}$:\\
Map of vacua:  the polynomial $Q(z)$ is
\begin{equation}\label{}\nonumber
  Q(z)=2z\prod^{N_{f}}_{p=1}\left(z^{2}-\widehat{z}^{2}_{p}\right).
\end{equation}
The match of the vacua
\begin{equation}\label{}
  {\cal} S(N_{c}, N_{f})\Leftrightarrow{\cal} S(N^{D}_{c}, N_{f}).
\end{equation}
The correlation function of $SO(2N_{c}+1)$ is equ'n (\ref{CLSOk4})
\begin{equation} \nonumber
\langle \cO_{0}+\rm{Pf}(\sigma)  \cO_{1}(\sigma^{2}_{a})\rangle_{g}=\sum_{p\in \mathcal{S}(N_{c}, N_{f})}\cO_{0}(\widehat{\sigma}^{2}_{p})\textit{H}(\widehat{\sigma}_{p})^{g-1}.
\end{equation}
The correlation function of $O_{+}(N-k+1)$ is
\begin{equation}\nonumber
\langle \cO_{0}+\tau\cdot\cO_{1}(\sigma^{2}_{a^{D}})\rangle_{g}=\sum_{\textit{p}^{D}\in \mathcal{S}(N^{D}_{c}, N_{f})}\cO_{0}(\widehat{\sigma}^{2}_{\textit{p}^{D}})\left(2H_{D}(\widehat{\sigma}_{\textit{p}^{D}})\right)^{g-1}.
\end{equation}
The match of the Hessian is given by
\begin{equation}\label{}\nonumber
  \frac{H(\widehat{\sigma}_{p})}{2H_{D}(\widehat{\sigma}_{\textit{p}^{D}})}=\frac{C}{C_{D}}2^{-4N^{D}_{C}-2}e^{i\pi \left(N_{c}N^{D}_{c}+\nu+N^{D}_{C}\right)}=1,
\end{equation}
where
\begin{equation}\label{}\nonumber
  C_{D}=2^{-4N^{D}_{C}-2}e^{i\pi \left(N_{c}N^{D}_{c}+\nu+N^{D}_{C}\right)}C.
\end{equation}
\end{itemize}

\section{$O_{-}(k)/O_{-}(N-k+1)$ in the mirror}\label{OOD}
The mirror of these gauge theories are parallel to the previous section, so we will only list the results.\\
\begin{itemize}
\item \noindent $O_{-}(2N_{c})\leftrightarrow O_{-}(2N^{D}_{c}+1)$, $N=2N_{f}$, $N^{D}_{c}=N_{f}-N_{c}$:\\
Map of vacua: $P(z)$ is
\begin{equation}\label{}\nonumber
  P(z)=2\prod^{N_{f}}_{p=1}\left(z^{2}-\widehat{z}^{2}_{p}\right).
\end{equation}
The ground state of the $O_{-}(2N_{c})$ theory is represented by the tuples
\begin{equation}\label{}\nonumber
  \left(\widehat{\sigma}_{p_{1}},\cdots, \widehat{\sigma}_{p_{N_{c}}}\right), \quad\quad p_{a}\in {\cal S}\left(N_{c}, N_{f}\right).
\end{equation}
These vacua do not intersect with orbifold $\mathbb{Z}_{2}$ fixed point, so we do not need to consider the twisted sector of this case. The ground states of the dual theory $O_{-}(2N^{D}_{c}+1)$ theory can be written in terms of $N^{D}_{c}$ length tuples
\begin{equation}\label{}\nonumber
  \left(\widehat{\sigma}_{p^{D}_{1}}, \cdots,\widehat{\sigma}_{p^{D}_{N^{D}_{c}}} \right), \quad\quad p^{D}_{a}\in {\cal S}\left(N^{D}_{c}, N_{f}\right).
\end{equation}
 The map of vacua is similar to the previous case that $p_{a}\leftrightarrow p^{D}_{a}$.\\
 The correlation function of $O_{-}(2N_{c})$ case is
 \begin{equation}\label{OEE-}
   \langle{\cal} O\rangle_{g}=\frac{1}{2}\langle{\cal} O\rangle_{g, SO(2N_{c})}=\frac{1}{2}\sum_{p\in \mathcal{S}(N_{c}, N_{f})}2\cO_{0}(\widehat{\sigma}^{2}_{p}){H}(\widehat{\sigma}_{p})^{g-1}=\sum_{p\in \mathcal{S}(N_{c}, N_{f})}\cO_{0}(\widehat{\sigma}^{2}_{p}){H}(\widehat{\sigma}_{p})^{g-1}.
 \end{equation}
The correlation function of $O_{-}(2N^{D}_{c}+1)$ case is
 \begin{equation}\label{OOE-}
   \langle{\cal} O\rangle_{g}=\frac{\left({\cal}{v}+3{\cal}{w}\right)^{g}}{2}\langle{\cal} O\rangle_{g, SO(2N^{D}_{c}+1)}=\sum_{p^{D}\in \mathcal{S}(N^{D}_{c}, N_{f})}\cO_{0}(\widehat{\sigma}^{2}_{p^{D}})\left(2H_{D}(\widehat{\sigma}_{p^{D}})\right)^{g-1}.
 \end{equation}
Here we have set ${\cal}{v}=-1$ and ${\cal}{w}=1$ from the sphere as well as the torus correlation function.\\

Again, the match of the correlation function can be transferred to the match of the Hessian, and one can find that
\begin{equation}\label{}\nonumber
  \frac{H(\widehat{\sigma}_{p})}{2H_{D}(\widehat{\sigma}_{p^{D}})}=\frac{C}{C_{D}}2^{-4N^{D}_{c}-1}e^{i\pi\left(N_{c}N^{D}_{c}+\nu\right)}=1,
\end{equation}
where
\begin{equation}\label{}\nonumber
  C_{D}=2^{-4N^{D}_{c}-1}e^{i\pi\left(N_{c}N^{D}_{c}+\nu\right)}C.
\end{equation}
\\
\item \noindent $O_{-}(2N_{c})\leftrightarrow O_{-}(2N^{D}_{c})$, $N=2N_{f}+1$, $N^{D}_{c}=N_{f}-N_{c}+1$:\\
The discussion of the map of vacua between the two dual theories is parallel to the case $SO(2N_{c})\leftrightarrow O_{+}(2N^{D}_{c})$. We simply list the result,
\begin{eqnarray}\nonumber
  \mathcal{S}(N_{c}, N_{f}) &\Leftrightarrow& \mathcal{S}(N^{D}_{c}-1, N_{f}) \\\nonumber
   \mathcal{S}(N_{c}-1, N_{f}) &\Leftrightarrow&  \mathcal{S}(N^{D}_{c}, N_{f}).
\end{eqnarray}
The orbifold twisted sectors have been projected out which also should be reflected in the correlation functions.\\
The correlation function of $O_{-}(2N_{c})$ is
\begin{eqnarray}\label{OEO-}\nonumber
 \langle \cO_{0}+\tau\cdot\cO_{1}(\sigma^{2}_{a})\rangle_{g}  &=& \frac{1}{2}\left(\sum_{p\in \mathcal{S}(N_{c}, N_{f})}2\cO_{0}(\widehat{\sigma}^{2}_{p}){H}(\widehat{\sigma}_{p})^{g-1}+{\cal}{v}^{g}\sum_{p\in \mathcal{S}(N_{c}-1, N_{f})}\cO_{0}(\widehat{\sigma}^{2}_{p}){H}(\widehat{\sigma}_{p},0)^{g-1}\right) \\
   &+&\frac{1}{2}\left[\left({\cal}{v}+3{\cal}{w}\right)^{g}-{\cal}{v}^{g}\right]\left( \sum_{p\in \mathcal{S}(N_{c}-1, N_{f})}\cO_{0}(\widehat{\sigma}^{2}_{p}){\mathcal{H}}(\widehat{\sigma}_{p},0)^{g-1}\right)\\\nonumber
   &=&\sum_{p\in \mathcal{S}(N_{c}, N_{f})}\cO_{0}(\widehat{\sigma}^{2}_{p}){H}(\widehat{\sigma}_{p})^{g-1}+\sum_{p\in \mathcal{S}(N_{c}-1, N_{f})}\cO_{0}(\widehat{\sigma}^{2}_{p})\left({2H}(\widehat{\sigma}_{p},0)\right)^{g-1},
\end{eqnarray}
where we have set ${\cal}{v}=-1$ and ${\cal}{w}=1$ for projecting the twisted sector. For vacua do not intersect with the orbifold fixed point, we do not need to consider the twisted sector. This explains our choice of ${\cal}{v}$ and ${\cal}{w}$ is different from the choice in \cite{Closset:2017vvl}. Furthermore, we have no overall minus sign mismatch of the correlation functions between two dual theories.\\
The correlation function of $O_{-}(2N^{D}_{c})$ is
\begin{eqnarray}\label{OEO-2}\nonumber
 \langle \cO_{0}&+&\tau\cdot\cO_{1}(\sigma^{2}_{a^{D}})\rangle_{g} \\\nonumber
  &=& \frac{1}{2}\left(\sum_{p^{D}\in \mathcal{S}(N^{D}_{c}, N_{f})}2\cO_{0}(\widehat{\sigma}^{2}_{p^{D}}){H_{D}}(\widehat{\sigma}_{p^{D}})^{g-1}+{\cal}{v}^{g}\sum_{p^{D}\in \mathcal{S}(N^{D}_{c}-1, N_{f})}\cO_{0}(\widehat{\sigma}^{2}_{p^{D}}){H_{D}}(\widehat{\sigma}_{p^{D}},0)^{g-1}\right) \\
   &+&\frac{1}{2}\left[\left({\cal}{v}+3{\cal}{w}\right)^{g}-{\cal}{v}^{g}\right]\left( \sum_{p^{D}\in \mathcal{S}(N^{D}_{c}-1, N_{f})}\cO_{0}(\widehat{\sigma}^{2}_{p^{D}}){H_{D}}(\widehat{\sigma}_{p^{D}},0)^{g-1}\right)\\\nonumber
   &=&\sum_{p^{D}\in \mathcal{S}(N^{D}_{c}, N_{f})}\cO_{0}(\widehat{\sigma}^{2}_{p^{D}}){H_{D}}(\widehat{\sigma}_{p^{D}})^{g-1}+\sum_{p^{D}\in \mathcal{S}(N^{D}_{c}-1, N_{f})}\cO_{0}(\widehat{\sigma}^{2}_{p^{D}})\left({2H_{D}}(\widehat{\sigma}_{p^{D}},0)\right)^{g-1},
\end{eqnarray}
where we again have set ${\cal}{v}=-1$ and ${\cal}{w}=1$ for projecting the twisted sector.\\

The match of the correlation functions can be tested by the match of the Hessian, and one can find
\begin{equation}\label{H21m1}\nonumber
  \frac{H(\widehat{\sigma}_{p_{a}})}{2H_{D}(\widehat{\sigma}_{p^{D}_{a^{D}}},0)}=\frac{2H(\widehat{\sigma}_{p_{a}},0)}{H_{D}(\widehat{\sigma}_{p^{D}_{a^{D}}})}=\frac{C}{C_{D}}2^{-4N^{D}_{c}+1}e^{i\pi \left(N_{c}N^{D}_{c}+\nu\right)}=1,
\end{equation}
where
\begin{equation}\label{}\nonumber
  C_{D}=C\cdot2^{-4N^{D}_{c}+1}e^{i\pi \left(N_{c}N^{D}_{c}+\nu\right)}.
\end{equation}
\\
\item\noindent $O_{-}(2N_{c}+1)\leftrightarrow O_{-}(2N^{D}_{c}+1)$, $N=2N_{f}+1$, $N^{D}_{c}=N_{f}-N_{c}$:\\
The map of vacua is
\begin{equation}\label{}\nonumber
  {\cal} S(N_{c}, N_{f})\Leftrightarrow{\cal} S(N^{D}_{c}, N_{f}).
\end{equation}
This map also include the twisted sector.\\

The correlation function of $O_{-}(2N_{c}+1)$ is
\begin{eqnarray}\label{ODD-}
  \langle \cO_{0}+\tau\cdot\cO_{1}(\sigma^{2}_{a})\rangle_{g} &=& \frac{\left({\cal}{v}+3{\cal}{w}\right)^{g}}{2}\sum_{\textit{p}\in \mathcal{S}(N_{c}, N_{f})}\cO_{0}(\widehat{\sigma}^{2}_{\textit{p}})H(\widehat{\sigma}_{\textit{p}})^{g-1} \\\nonumber
   &=&  2\sum_{\textit{p}\in \mathcal{S}(N_{c}, N_{f})}\cO_{0}(\widehat{\sigma}^{2}_{\textit{p}})\left(4H(\widehat{\sigma}_{\textit{p}})\right)^{g-1}.
\end{eqnarray}
We have set ${\cal}{v}={\cal}{w}=1$ in above.\\
The correlation function of $O_{-}(2N^{D}_{c}+1)$ is similar:
\begin{eqnarray}\label{ODD-2}
  \langle \cO_{0}+\tau\cdot\cO_{1}(\sigma^{2}_{a^{D}})\rangle_{g}&=&\frac{\left({\cal}{v}+3{\cal}{w}\right)^{g}}{2}\sum_{\textit{p}^{D}\in \mathcal{S}(N^{D}_{c}, N_{f})}\cO_{0}(\widehat{\sigma}^{2}_{\textit{p}^{D}})H_{D}(\widehat{\sigma}_{\textit{p}^{D}})^{g-1} \\\nonumber
   &=& 2\sum_{\textit{p}^{D}\in \mathcal{S}(N^{D}_{c}, N_{f})}\left(\cO_{0}(\widehat{\sigma}^{2}_{\textit{p}^{D}})\right)\left(4H_{D}(\widehat{\sigma}_{\textit{p}^{D}})\right)^{g-1}.
\end{eqnarray}
The match of Hessian is
\begin{equation}\label{}\nonumber
  \frac{H(\widehat{\sigma}_{\textit{p}})}{H_{D}(\widehat{\sigma}_{\textit{p}^{D}})}=\frac{C}{C_{D}}2^{-4N^{D}_{c}-1}e^{i\pi\left( N_{c}N^{D}_{c}+N^{D}_{c}+\nu\right)}=1,
\end{equation}
where we impose
\begin{equation}\label{}\nonumber
  C_{D}=2^{-4N^{D}_{c}-1}e^{i\pi\left( N_{c}N^{D}_{c}+N^{D}_{c}+\nu\right)}C.
\end{equation}
\end{itemize}

\section*{Acknowledgement}
We would like to thank Cyril Closset, Enno Ke{\ss}ler, Daniel S.Park and Eric Sharpe for useful discussions and comments.

\end{document}